\documentclass[twocolumn]{aastex6}

\usepackage{amsmath}
\makeatletter
\def\env@matrix{\hskip -\arraycolsep 
  \let\@ifnextchar\new@ifnextchar
  \array{*{\c@MaxMatrixCols}c}}
\makeatother
\usepackage{graphicx,psfrag,epsf}
\usepackage{enumerate}
\usepackage{placeins}
\usepackage{xspace}
\usepackage{subfigure}
\usepackage{booktabs}
\PassOptionsToPackage{hyphens}{url}\usepackage{hyperref}
\usepackage{bbm}

\newcommand{\Kep}{\emph{Kepler}\xspace}

\newcommand{\Rearth}{R$_\oplus$\xspace}
\newcommand{\Mearth}{M$_\oplus$\xspace}

\slugcomment{Submitted to the Astrophysical Journal}
\shorttitle{Nonparametric Exoplanet Mass-Radius Relation}
\shortauthors{Ning, Wolfgang, \& Ghosh}

\begin{document}

\title{Predicting Exoplanets Mass and Radius: A Nonparametric Approach}

\author{Bo Ning\altaffilmark{1}, Angie Wolfgang\altaffilmark{2,3,4}, and Sujit Ghosh\altaffilmark{5}}
\altaffiltext{1}{Department of Statistics and Data Science, Yale University, 24 Hillhouse Ave, New Haven, CT 06511}
\altaffiltext{2}{Department of Astronomy and Astrophysics, Pennsylvania State University,
    University Park, PA 16802}
\altaffiltext{3}{Center for Exoplanets and Habitable Worlds, 525 Davey Laboratory, The Pennsylvania State University, University Park, PA, 16802, USA.}
\altaffiltext{4}{NSF Astronomy \& Astrophysics Postdoctoral Fellow}
\altaffiltext{5}{Department of Statistcs, North Carolina State University, 2311 Stinson Dr., Raleigh, NC 27695}
\email{bo.ning@yale.edu; akw5014@psu.edu; sujit.ghosh@ncsu.edu}

\begin{abstract}

A fundamental endeavor in exoplanetary research is to characterize the bulk compositions of planets via measurements of their masses and radii.  
With future sample sizes of hundreds of planets to come from TESS and PLATO, we develop a statistical method that can flexibly yet robustly characterize these compositions empirically, via the exoplanet M-R relation. Although the M-R relation has been explored in many prior works, they mostly use a power-law model, with assumptions that are not flexible enough to capture important features in current and future M-R diagrams. To address these shortcomings, a nonparametric approach is developed using a sequence of Bernstein polynomials.  We demonstrate the benefit of taking the nonparametric approach by benchmarking our findings with previous work and showing that a power-law can only reasonably describe the M-R relation of the smallest planets and that the intrinsic scatter can change non-monotonically with different values of a radius. We then apply this method to a larger dataset, consisting of all the Kepler observations in the NASA Exoplanet Archive. Our nonparametric approach provides a tool to estimate the M-R relation by incorporating heteroskedastic measurement errors into the model. As more observations will be obtained in the near future, this approach can be used with the provided \textsf{R} code to analyze a larger dataset for a better understanding of the M-R relation. 

\end{abstract}


\section{Introduction}
\label{sec:intro}

An exoplanet's mass and radius can constrain its bulk composition when models of planetary internal structures are fit to these observed properties (e.g. \citealt{Val07,Sea07,For07a,Rog11,Lop14}).  Measuring many planets' masses and radii can therefore illuminate: 1) the range of exoplanet compositions produced by planet formation and evolution processes; and 2) how these compositions trend with planet and stellar properties of interest, such as planet mass, orbital period, host star mass, and host star metallicity.  The existence and nature of these trends, along with the amount of astrophysical scatter around them, provides observational constraints on planet formation theory (e.g. \citealt{Ida10,Lee15,Daw15,Lop16,Owen17}).

The population distributions of these mass and radius measurements therefore contain valuable information about the range of planetary compositions. To obtain the population distributions is particularly necessary for the thousands of super-Earth- and sub-Neptune-sized planets discovered by the \Kep mission \citep{Cou16}, as this population is not represented in our own Solar System.

Initially population studies of small, low-mass planets focused on the marginal mass \citep{How10,May11} or marginal radius \citep{You11,How12,Fre13,Pet13,Dre13} distributions; this was driven in part by the difficulty in obtaining a sample of planets which both transited their host stars and were suitable to precision radial velocity (RV) follow-up.  
This changed once the \Kep follow-up program published mass estimates for their chosen subset of \Kep candidates \citep{Mar14}. {Recent analysis of \Kep planets' transit timing variations have also produced robust mass determinations for dozens of multiple-planet systems \citep{Jon16,Mac16,Mil16,Had17}.}
However, mass measurements for {the majority of} individual \Kep planet candidates are {still} unavailable.  {As a result,}
the mass-radius (M-R) relations estimated based on the existing mass and radius measurements provides a useful tool for 
astronomers to predict masses based on observed radii for those planets {which} only have radius measurements.

\subsection{Previous work on mass-radius relations} \label{sec:litM-R}

There have been several M-R relations proposed in the exoplanet literature; all of them assume that exoplanet masses and radii follow a power-law:

\begin{itemize}
\item \citet{Lis11} fit a relation to Earth and Saturn and obtained $M=R^{2.06}$, where $M$ and $R$ are in Earth units.  
\item \citet{WuY13} calculated the masses of 22 planet pairs based on the amplitudes of their sinusoidal transit timing variations (TTVs) and found that $M=3R$. 
\item \citet{Wei14} used the 42 planets chosen by the \Kep team to be followed up with radial velocity measurements \citep{Mar14} and found the power law is $M=2.69R^{0.93}$ for planets with $1.5<R<4$ \Rearth.
\item \citet{Had14} revisited the M-R relation for 56 low-eccentricity TTV planets and fit 
$M=14.9 (R/3R_\oplus)^{0.65}$ to them.
\item \citet{Wol16} (hereafter denoted as WRF16) selected planets with radii $<8$ \Rearth from the NASA Exoplanet Archive (up to the date 1/30/2015) and found two different power-law relations: 
$M = 2.7R^{1.3}$ for $<4$ \Rearth planets and $M = 1.6R^{1.8}$ for $<8$\Rearth planets.
Unlike the earlier results, which used basic least squares regression, 
WRF16 built a hierarchical Bayesian model that incorporated observational measurement uncertainty, the intrinsic, astrophysical scatter of planet masses, and non-detections and upper limits.
\item \citet{Mil17} fit power laws to various subsets of exoplanet masses and radii, testing for differences between RV and TTV M-R relations in two different period bins.
\item \citet{Chen17} built on the hierarchical Bayesian model from WRF16 to fit a continuous broken power-law model across a much larger range of masses and radii.
They found that a four-segment M-R relation best described their data, with Terran, Neptunian, Jovian, and Stellar regions. They fit a different power-law and intrinsic scatter to each segment.
\end{itemize}

\subsection{Moving to a nonparametric approach} \label{sec:whynonparam}

The power-law model is simple to fit to exoplanet masses and radii and its parameters are easy to interpret. However, there are several reasons to move beyond it.  First, {we already know that} a single power-law relation will not be able to describe the M-R relation for the entire population, as degeneracy pressure {causes} the radius-mass relation to flatten out around a Jupiter mass.  
\citet{Chen17} addressed this by assuming a broken power-law with multiple segments.  
As soon as one expands the space of models beyond those with a handful of parameters, it is useful to consider more flexible nonparametric models as well\footnote{We note that ``nonparametric'' is a bit of a misnomer: nonparametric models still do have parameters.  
In our case, these parameters are the degree $d$ and the weights $w_{kl}$ in Eqn \ref{eq:bernMod-D}.  
What makes nonparametric models different than parametric ones is that {the dimension of the parameters} (in this case $d$) is allowed to vary with sample size{, which allows} for much greater flexibility to adapt to complex shapes.}. Indeed, there is no particular reason to believe that exoplanet masses and radii should follow power-laws as a population, since the processes that dominate planet formation for small rocky planets are different than those that dominate the formation of gas giants. Furthermore, we do not expect the distribution of masses at a given radius to be Gaussian, as was assumed by WRF16 and \citet{Chen17}, or even symmetric. 
There is already evidence from the hierarchical modeling checking performed by WRF16 that a Gaussian distribution does not completely reproduce the observed exoplanet masses and radii.

Moving forward, we adopt a nonparametric approach that allows us to relax these assumptions.  
There are many benefits of using nonparametric models, including:

\begin{enumerate}

\item {They can take on variety of shapes to fit the data, which can be advantageous for making predictions that are less model-dependent.  Therefore, there is no} need to impose abrupt breaks for modeling M-R relations {across a wide range of sizes}.

\item They can model the joint distribution of mass and radius, 
and thus provide a {self-consistent} way to predict {both} mass from radius and radius from mass over the entire exoplanet mass and radius range.  

\item Eventually we want to understand masses and radii as a function of many other star and planet properties like stellar mass, metallicity, orbital period, planet multiplicity, even eccentricity. We do not have clear guidance from theory about the functional form of the dependence in these additional dimensions, and so we should not expect the mass-radius-X distribution to follow a power law. 
\end{enumerate}

\section{The Model}
\label{sec:model}

There are at least two ways to model the exoplanet mass-radius relation.  The first is to approach it as a regression problem.  When one performs regression, such as using ordinary least squares (OLS) to fit a line to data, the goal is to illuminate the relationship between an independent variable and another quantity that depends on it---the dependent variable.  
For exoplanets, we are indeed concerned with how mass and radius are related, as it illuminates how planet compositions change as a function of the planet's size or mass.  However, it is not so clear which is the independent variable and which is the dependent variable.  From a theoretical point of view, mass is the more fundamental property and so should be the independent variable\footnote{That said, some physical processes which affect exoplanet compositions, such as photoevaporation, depend on both mass and radius.}.  
On the other hand, the small fraction of \Kep planets that only have mass measurements creates a practical need to convert \Kep radii to masses (or in other words, to predict masses from radii), which requires radius to be the independent variable.  

Moreover, there are non-negligible measurement uncertainties associated with both mass and radius.  
OLS regression, which ignores the uncertainties on the independent variable, will consistently underestimate the slope of the regression line when there are uncertainties in both variables (see \citealt{Iso90}).  This occurs because OLS only minimizes the distance between the points and the line in the vertical direction, which over-corrects for the additional vertical distance introduced by uncertainties in the dependent variable (for lines with slope $\ne 0$, the horizontal scatter produced by uncertainties in the dependent variable will also produce some additional vertical scatter around the line).  
As a result, the OLS line which fits mass to radius will not be the inverse of the line which fits radius to mass. 
This is a problem if our goal is to estimate the underlying fundamental relationship between radius and mass.

As our goal in this paper is to provide mass predictions from measured radii as well as radius predictions from mass, we decide to approach the problem differently. 
WRF16 already showed that a simple regression line, i.e. a one-to-one function which deterministically maps radius to mass, is an insufficient description of the existing data: there exists an intrinsic, astrophysical scatter in the mass-radius relation.  Therefore, we also want this approach to allow for a distribution of masses at any one radius.  We can express this distribution as $g(m|r)$, the conditional probability distribution of mass given radius, i.e. a function describing the probability of a planet having a certain mass when its radius takes a specific value\footnote{WRF16 modeled $g(m|r)$ as a normal distribution but provided some evidence that this assumption was too restrictive; this evidence is partly what has motivated us to adopt a nonparametric approach in this paper.}. Through the definition of conditional probability, every conditional probability distribution can be expressed as a ratio of the joint probability distribution $g(m,r)$ to the marginal probability distribution of the variable on which the conditioning occurs ($g(r)$):
\begin{equation} \label{eqn:cond}
g(m|r) = \frac{g(m,r)}{g(r)} = \frac{g(m,r)}{\int g(m,r)\text{d}m}
\end{equation}
Likewise,
\begin{equation} \label{eqn:cond-rm}
g(r|m) = \frac{g(m,r)}{g(m)} = \frac{g(m,r)}{\int g(m,r)\text{d}r}
\end{equation}
Therefore, to get both conditional distributions we need only to model the joint distribution: the probability of a planet having a certain mass \emph{and} a certain radius.  It is this joint distribution that we define in \S \ref{sec:ourmod} and fit to our data in \S \ref{sec:modinfer}.  For the rest of the paper, we will use $g(m,r)$ to discuss general properties of joint mass, radius distributions.  We will use $f(m,r)$ to refer to our specific formulation of the joint mass, radius distribution as given by Eqn \ref{eq:bernMod-D}.


\subsection{The Bernstein polynomials model} \label{sec:ourmod}

Our nonparametric model for the joint probability distribution of mass and radius is a bivariate distribution that consists of a tensor product of sequence of {location-scaled and transformed} beta densities which can also be viewed as Bernstein polynomials.
Each marginal distribution can be expressed as a linear combination of Bernstein polynomials (BPs), leading to mixture of beta densities.  When normalized, BPs have the same functional form as beta distributions (see Eqn \ref{eq:beta}). 
For a visual representation of Bernstein Polynomials, see Figure \ref{fig:BPterms}, where we plot BPs as a function of their degree (denoted by $k$ or $l$ in Eqn \ref{eq:bernMod-D}).  

{The properties of BPs and a discussion of their advantages over other choices for basis functions are described in Appendix \ref{sec:BPprop}.  In addition, we note that we do not use Gaussian process regression, which is a popular nonparametric prediction method in exoplanet science, for a number of reasons.  First, a joint density estimation approach is more appropriate for our purposes than a regression approach, as argued above. Second, the marginal distribution of the M-R is not in general normally distributed, as we demonstrate in Section \ref{sec:jointresult}.  Third, fitting a mixture of Gaussians to describe the joint density would involve three free parameters per component (mixture weight, mean, and variance) rather than the one free parameter per component that we use here (the mixture weight).  Fourth, a mixture of Gaussians produces an identifiability problem, wherein different components can be interchangeable, that is not present in the Bernstein polynomial model due to the ordered nature of the different terms (see Eqn \ref{eq:bernMod-D} and the figures in Appendix \ref{sec:BPprop} for how the Bernstein polynomial terms, i.e. the different $\beta_{k}$ for a given $d$, are distinguishable from each other).}

Before introducing our model, we first define some mathematical notations that will be used throughout the paper.
Let $M_i^{\text{obs}}$ and $R_i^{\text{obs}}$ be the reported
mass and radius measurements for $i$-th planet {in our dataset},
and $\sigma_{M_i}^{\text{obs}}$ and $\sigma_{R_i}^{\text{obs}}$ be the reported standard deviations for their measurement errors.
We denote $M_i$ and $R_i$ as the true mass and radius for $i$-th planet that would have been observed if there were no measurement errors.
We denote their joint distribution as $f(m, r)$,
and the M-R relation as the conditional distribution $f(m| r)$.
We let $\boldsymbol M^\text{obs}$ stand for a vector containing the observations $(M_1^\text{obs}, \dots, M_n^\text{obs})$; similarly, $\boldsymbol R^\text{obs}$, $\boldsymbol \sigma_{M}^\text{obs}$ and $\boldsymbol \sigma_{R}^\text{obs}$ stand for the set of their respective observations.
We let $\mathcal{N}(\mu, \sigma)$ stand for a normal distribution with 
mean $\mu$ and standard deviation $\sigma$, and denote the probability density of a Beta distribution with shape parameters $i$ and $d-i+1$ by
\begin{equation}
\label{eq:beta}
\beta_{id}(a) = d {{d-1} \choose {i-1}} a^{i-1} (1-a)^{d-i}, 
\end{equation} 
where $0\leq a\leq 1$, and $d>0$ is an integer that relates to the degree of a Bernstein polynomial.

Our hierarchical model can be written as follows:
\begin{align}
  &M^{\text{obs}}_i \overset{ind}{\sim} 
  \mathcal{N}(M_i, \sigma^{{\text{obs}}}_{M_i}),
  \label{eq:bernMod-A} \\     
  &R^{\text{obs}}_i \overset{ind}{\sim} 
  \mathcal{N}(R_i, \sigma^{{\text{obs}}}_{R_i}),
  \label{eq:bernMod-B} \\
  &(M_i, R_i) \overset{iid}{\sim} f(m, r| \boldsymbol w, d, d'),
       \label{eq:bernMod-C}
\end{align}
where
the symbol
$\overset{ind}{\sim}$ stands for ``independently distributed as", and the symbol
$\overset{iid}{\sim}$ stands for ``identically and independently distributed as".  
Let
$\boldsymbol w = (w_{11}, \dots, w_{dd'}$) be a set of weights {which describe how much each corresponding term} in the following series contribute to the overall joint distribution $f(m,r)$:
\begin{equation}
    f(m, r| \boldsymbol w, d, d')  
    = \sum_{k=1}^{d}
    \sum_{l=1}^{d'} w_{kl} 
		\frac{\beta_{kd}(\frac{m - \underline{M}}{\overline{M} - \underline{M}})}{\overline{M} - \underline{M}}
		\frac{\beta_{ld'}(\frac{r - \underline{R}}{\overline{R} - \underline{R}})}{\overline{R} - \underline{R}}.
        \label{eq:bernMod-D}
\end{equation}
To ensure that Eqn \ref{eq:bernMod-D} remains a probability distribution{, i.e. that} it integrates to 1, we impose the constraint 
$\sum_{k=1}^d \sum_{l=1}^{d'} w_{kl} = 1$ and that $w_{kl}\geq 0$ {for} all values of $k$ and $l$.  
The notations
$\underline{M}$, $\overline{M}$, $\underline{R}$ and $\overline{R}$ are used to denote the lower and upper bounds, respectively, for mass and radius.
The values of the upper and lower bounds can be determined by a set of observed values of masses and radii; for example, 
one could choose the lower mass bound to be the minimum ($M_i^{\text{obs}}-\sigma_{M_i}^{\text{obs}}/n$) in the dataset,
and the upper mass bound to be the maximum 
($M_i^{\text{obs}} + \sigma_{M_i}^{\text{obs}}/n$).
The bounds could also be set as the minimum and maximum mass and radius theoretically expected for an exoplanet.  
{For regions with no observations, the BP model reverts to the overall mean of the whole function;}
this happens when the values of $\underline{M}$, $\overline{M}$, $\underline{R}$ and $\overline{R}$ are chosen to be far away from the nearest observations {(see discussion in \S \ref{sec:bounds})}.

Note that Eqns \ref{eq:bernMod-A}--\ref{eq:bernMod-C} form a multi-level model, in that the measurement process is modeled as a separate random process from the underlying true distribution of exoplanet masses and radii.  It therefore has a similar hierarchical structure as the hierarchical Bayesian models defined in WRF16 and \citet{Chen17}.  

\subsection{Model inference} \label{sec:modinfer}

Although the model may look complex at the first glance, 
the inference is relatively straightforward.
There are only two types of unknown parameters in the model,
the weights $\boldsymbol w$ and the degrees $d$ and $d'$.  
Once $d$ and $d'$ are determined {using the method we will describe below,} the two parameters {$i$ and $d-i+1$} in each beta distribution {(see Eqn \ref{eq:beta})} are also determined. 
This is advantageous compared to other mixture models; for example, a mixture of Gaussians would require that each mean and standard deviation of component distribution be estimated by the data.  
Moreover, it avoids the common identifiability issues {created by label switching within the mixture} of Gaussians. Another {advantage} of our mixture of beta {distributions} is that parameter space (which is a simplex) is a bounded convex set{,} which {helps to guarantee} the existence of the optimally estimated parameters. Here we take a maximum likelihood approach to estimate the parameters by explicitly deriving the likelihood using one-dimensional numerical integration.

The likelihood function of the model \ref{eq:bernMod-A}--\ref{eq:bernMod-C}
can be written as, 
\begingroup
\allowdisplaybreaks
\begin{align}
&  L(\boldsymbol w, d, d'| \boldsymbol M^{\text{obs}}, 
   \boldsymbol R^{\text{obs}}, 
   \boldsymbol \sigma^{{\text{obs}}}_{M}, 
   \boldsymbol \sigma^{{\text{obs}}}_{R})
   \nonumber\\
& =
    \int_{\underline{M}}^{ \overline{M}} 
    \int_{\underline{R}}^{ \overline{R}} 
	f(\boldsymbol M^{\text{obs}}, 
	  \boldsymbol R^{\text{obs}}, m, r | 
		\boldsymbol w, d, d',
		\boldsymbol \sigma^{{\text{obs}}}_M, 
		\boldsymbol \sigma^{{\text{obs}}}_R)
	\ \text{d}r \text{d}m 
	\nonumber\\
 & = 
	\prod_{i=1}^n     
	\int_{\underline{M}}^{ \overline{M}} 
    \int_{\underline{R}}^{ \overline{R}} 
	f(M^{\text{obs}}_i| m, \sigma^{{\text{obs}}}_{M_i}) 
    f(R^{\text{obs}}_i| r,  \sigma^{{\text{obs}}}_{R_i})
    \nonumber\\
& \quad \times    
    f(m, r| \boldsymbol w, d, d') 
     \ \text{d}r \text{d}m
     \nonumber\\
 & = 
	\prod_{i=1}^n \sum_{k=1}^{d} \sum_{l=1}^{d'} w_{kl}
	\int_{\underline{M}}^{ \overline{M}}  
    \frac{1}{\sigma_{M_i}^{\text{obs}}}
	\mathcal{N} \Big (
    	\frac{M_i^{\text{obs}} - m}{\sigma_{M_i}^{\text{obs}}} 
    \Big )
	\frac{\beta_{kd}(\frac{m - \underline{M}}
    {\overline{M} - \underline{M}})}{\overline{M} - \underline{M}} 
				\ \text{d}m 
	\nonumber\\
& \quad \times 
   \int_{\underline{R}}^{ \overline{R}} 
   \frac{1}{\sigma_{R_i}^{\text{obs}}}
   \mathcal{N} \Big(\frac{R_i^{\text{obs}} - r}{\sigma_{R_i}^{\text{obs}}} \Big)
   \frac{\beta_{ld'}(\frac{r - \underline{R}}{\overline{R} - \underline{R}})} 
   {\overline{R}-\underline{R}} \ \text{d}r.
   \label{eqn:likelihood}
\end{align}
\endgroup

The two integral terms are essentially two constants  
and can be calculated numerically by using 
{\tt integrate} function in $\mathsf{R}$ or by any other Gaussian quadrature method available for one-dimensional numerical integration.
Therefore, we denote
\begin{equation*}
\begin{split}
c_{kl,i} & \equiv
	\int_{\underline{M}}^{ \overline{M}} 
	\frac{1}{\sigma_{M_i}^{\text{obs}}}
	\mathcal{N} \Big (\frac{M_i^{\text{obs}} - m}
    					{\sigma_{M_i}^{\text{obs}}} 
    \Big )
    \frac{\beta_{kd}(\frac{m - \underline{M}}
    {\overline{M} - \underline{M}})}{\overline{M} - \underline{M}} 
	\ \text{d}m \\
	& \quad \times
    \int_{\underline{R}}^{\overline{R}} 
	\frac{1}{\sigma_{R_i}^{\text{obs}}}
	\mathcal{N} \Big(\frac{R_i^{\text{obs}} - r}
    {\sigma_{R_i}^{\text{obs}}} \Big)
    \frac{\beta_{ld'}(\frac{r - \underline{R}}
    {\overline{R} - \underline{R}})}{\overline{R}-\underline{R}}
    \ \text{d}r,
\end{split}
\end{equation*}
and $\boldsymbol c_i = (c_{11,i}, \dots, c_{dd',i})$
to simplify the expression.

Then for given values $d$ and $d'$, we find an estimator of $\boldsymbol w$ 
which maximizes the log-likelihood,
\begin{equation}
\begin{split}
\label{eqn:cp}
 \text{maximize:}
&	\quad  \log L
	= \sum_{i=1}^n \log(\boldsymbol c_i^T \boldsymbol w), \\
  \text{subject to:} 
&	\quad \sum_{k=1}^d \sum_{l=1}^{d'} 
w_{kl} = 1, w_{kl} \geq 0,\\
& \quad \text{for all } k = 1, \dots, d,\ 
l = 1, \dots, d'.
\end{split}   
\end{equation}
The above problem is a convex optimization problem which can be readily solved using any standard numerical optimization methods. For our applications, we have used the $\mathsf{R}$ package {\tt Rsolnp} to solve the above optimization problem.

Now we discuss how we choose $d$ and $d'${, which act} as tuning parameters relative to the smoothness of the {underlying} density function. 
We first choose a set of candidate values for $d$ and $d'$, 
{such as} 2, 3, $\dots$, $n/\log(n)$, 
where $n$ is the sample size (i.e. number of exoplanets). 
We use 10-fold cross validation to choose the optimal value of degrees $d$ and $d'$
from those candidate values.
To conduct the 10-fold cross validation,
we separate the dataset randomly into 10 disjoint subsets 
with equal sizes.
Then we leave {out} the $s$-th subset, denoted by $D_s$ {(}$s = 1, \dots, 10${)} and use the remaining 9 subsets of data to estimate the parameters{.  Doing this for each $D_s$ in turn results in} 10 estimated sets of weights, 
$\hat{\boldsymbol w}^{(s)}$ based all observations $i\notin D_s$.
We {plug in} each $\hat{\boldsymbol w}^{(s)}$
along with the {corresponding} data that {was} used to estimate {each set of} weights
to obtain an estimated value for the log-likelihood.
Mathematically, we are calculating the following quantity,
\begin{equation*}
	\log L^{\text{pred}} 
	= 
	\sum_{s=1}^{10} \sum_{i\notin D_s}\log(\boldsymbol c_i^T \hat{\boldsymbol w}^{(s)})
\end{equation*}
for different possible values of $d$ and $d'${.  The optimum degrees for the BP model are then the $d$ and $d'$ which give} the largest value of $-\log L^{\text{pred}}$.

Besides using the cross-validation method, 
other popular methods such as AIC (Akaike information criterion)
and BIC (Bayesian information criterion) are also possible when 
data samples are abundant.  
When the sample size is relatively small, both methods will under-select the degrees,
as in the example provided in \citet{Tur14}. 
This is the reason why we choose to use the 10-fold cross validation method.

The model may require $d$ and $d'$ {to be large},
and thus the number of weights {which need} to be estimated is also large.
In such a situation, 
one may be concerned {whether} we have enough data to estimate the weights{.  Fortunately}, a majority of the estimated weights turns out to be numerically {zero, due to the fact that parameter space is a simplex and therefore convex.  This drastically reduces the number of effective parameters in our model.  For example, mass predictions computed using the largest 25 weights of the fit performed in \S \ref{sec:jointresult} are indistinguishable from predictions produced using the full set of weights, to the 3rd or 4th decimal in log($M$).}

Once we obtained $\hat d$, $\hat d'$ and $\hat{\boldsymbol w}$, 
the estimated values of $d$, $d'$ and $\boldsymbol w$,
the M-R relation can be derived {following Eqn} \ref{eqn:cond}: 
\begin{align}
    f(m|r, \hat{\boldsymbol w}, \hat d, \hat{d}') 
    & = 
    \frac{f(m, r| \hat{\boldsymbol w}, \hat d,
    \hat{d}') }{\int_{\underline{M}}^{\overline{M}} 
    f(m, r| \hat{\boldsymbol w}, \hat d, \hat{d}') \text{d}m} 
    \nonumber \\
    & = 
    \frac{f(m, r| \hat{\boldsymbol w}, \hat d, \hat d') }{f(r| \hat{\boldsymbol w}, \hat d, \hat d')},
    \label{equ:condMR}
\end{align}
where 
\begin{equation*}
	f(r| \hat{\boldsymbol w}, \hat d, \hat d')
	=
	\sum_{k=1}^{\hat d} \sum_{l=1}^{\hat d'}
	\hat w_{kl} \frac{
	\beta_{l\hat d'}\Big( \frac{r - \underline{R}}{\overline{R}-\underline{R}}  \Big)}
  {(\overline{R} - \underline{R})}.
\end{equation*}

From Eqn \ref{equ:condMR}, mean, variance and prediction intervals
can be obtained {analytically}. 

As noted in \S \ref{sec:ourmod}, we could also estimate our nonparametric model from a Bayesian framework
by assigning priors to the parameters $\boldsymbol w$, $d$ and $d'$.  
For example we can assign a Dirichlet prior to the weights as follows:
\begin{equation*}
   \pi(w_{11}, w_{12} \dots, w_{dd'}) \sim \text{Dir}(\alpha_{11}, \alpha_{12},\dots, \alpha_{dd'}),
\end{equation*}
where 
$\sum_{k=1}^d \sum_{l=1}^{d'} {\alpha_{kl}} = 1$,
and assign a Poisson or uniform prior for $d$
and $d'$ respectively.
The inference can be carried out using a
Markov chain Monte Carlo algorithm (MCMC) (i.e. \citealt{Pet99a, Pet99b}). 
However, {initial investigations into this approach indicated that} the MCMC algorithm {would take a} much longer time to run{.  As a result, we employ a maximum likelihood method rather than a Bayesian method} in this paper.

\section{Data}
\label{sec:data}

We apply the Bernstein polynomial model to obtain the M-R relations from two datasets.  
The first dataset is taken from WRF16 to enable a direct comparison between their parametric mass-radius relation and our nonparametric one (this comparison is illustrated in Figure \ref{fig:MRrelations}).  
Specifically, we use their extended radial velocity dataset, denoted in WRF16 as ``RV only, $<8$\Rearth" and which consists of all planets in their Table 2 except those labeled ``c".  In this work we also exclude the planets whose mass measurements are only upper limits, as the \textsf{R} package we use to find the maximum likelihood estimates of the weights $\boldsymbol w$ does not allow censored data.  The results from this benchmark dataset ($N=60$) are presented in \S \ref{sec:bench}.

The second dataset consists of all planets with an assigned \Kep name, whose mass measurements originated from either radial velocities (RV) or from N-body dynamical fits to transit timing variations (TTVs).  {We note that TTV planets could have astrophysically different densities;
however, we see that the inclusion of high-quality TTV masses in the current manuscript as a reasonable decision, both because prior work has made the same decision (\citealp{Wei14} and WRF16) and because a dataset that contains both TTV and RV masses better represents the full range of densities that an M-R relation would need to reproduce.} 
We obtained this information from the NASA Exoplanet Archive \citep{Ake13} on June 5, 2017; for this dataset we also excluded planets with only upper limits reported on their masses.  With these restrictions, our \Kep-only dataset consists of 127 planets with robust mass measurements; the results from this science dataset are presented in \S \ref{sec:jointresult}.

\subsection{Discussion of selection effects} \label{sec:seleffect}

We choose to restrict our science dataset (results presented in \S \ref{sec:jointresult}) 
to \Kep planets in an effort to minimize the influence of survey selection effects.  Selection effects manifest in two ways for the mass-radius relation:
\begin{enumerate}
\item The probability of detecting a planet is a non-constant function of either their mass (for RV detection) or radius (for transit detection), with smaller or less massive planets being more difficult to detect.  
\item The probability that a known planet has its radius (for RV detections) or its mass (for transit detections) measured by follow-up observations is a complicated, often unpublished function of the planet's discovery mass/radius, the predicted radius/mass, and the host star's brightness, spectral type, activity indicators, sky position, etc.  
\end{enumerate}

These two selection effects impact the inferred mass-radius relation in different ways.  The first affects the relative amount of data at large vs. small radii (or at high vs. low mass).  Correcting for this effect becomes important when one's scientific goal is a joint mass, radius probability distribution $g(m,r)$ that reflects the underlying, true distribution of exoplanet masses and radii.  For example, if one tried to use the data from the Exoplanet Archive as-is with no correction for detection completeness, they would incorrectly conclude that $\sim 1.1$R$_\text{Jupiter}$ and $\sim 1$M$_\text{Jupiter}$ is the most probable mass and radius for an exoplanet, as more Jupiter-sized planets have had their masses measured than smaller planets. 
We already know that this potential conclusion to be incorrect: the numerous occurrence rate studies which use \Kep data to correct for variable detection efficiency across the full range of planet radii (e.g. most recently \citealt{FoM14,Ful17,Hsu18}) have shown that smaller planets are much more prevalent than Jupiter-sized ones.  Therefore, if one wishes to understand the probability of an exoplanet existing with a specific mass \emph{and} radius (i.e. characterize $g(m,r)$), they must account for the different surveys' detection completeness.

Fortunately, this is not the goal of this work.  We are scientifically interested in $g(m|r)$, the conditional distribution which gives us the M-R relation.  {In going from $g(m,r)$ to $g(m|r)$, we marginalize over $r$, which divides out the completeness correction in that dimension.  With this correction disappearing from the equation for $g(m|r)$, we can safely ignore the} effect of the selection bias due to detectability of different radii planet.
(see Eqn \ref{eqn:cond}).  {This leaves only the second concern for our mass predictions: how transiting planets are selected for RV follow-up.}  

{To address this concern}, one needs to know how planets were chosen for follow-up.  
To date, no RV follow-up group has published their selection function in a quantitative way that would allow us to incorporate it into an analysis like this.  Furthermore, given that the target list often evolves as RV data are acquired, this selection function is probably intractable to calculate for the current dataset.  Progress is being made toward this end by designing RV follow-up campaigns that have definable selection functions (i.e., \citealp{Burt18} and \citealp{Mont18}), but for the current dataset this remains practically out of reach.
No other papers on the M-R relation has incorporated {corrections for} completeness or selection effects into their analysis.  
{We follow this precedent, acknowledging that is a clear area for future work, and choose to focus instead on the novel} development of a nonparametric analysis.

\section{Benchmarking to WRF16}
\label{sec:bench}

\begin{figure*}
\centering
   \includegraphics[width=9cm]{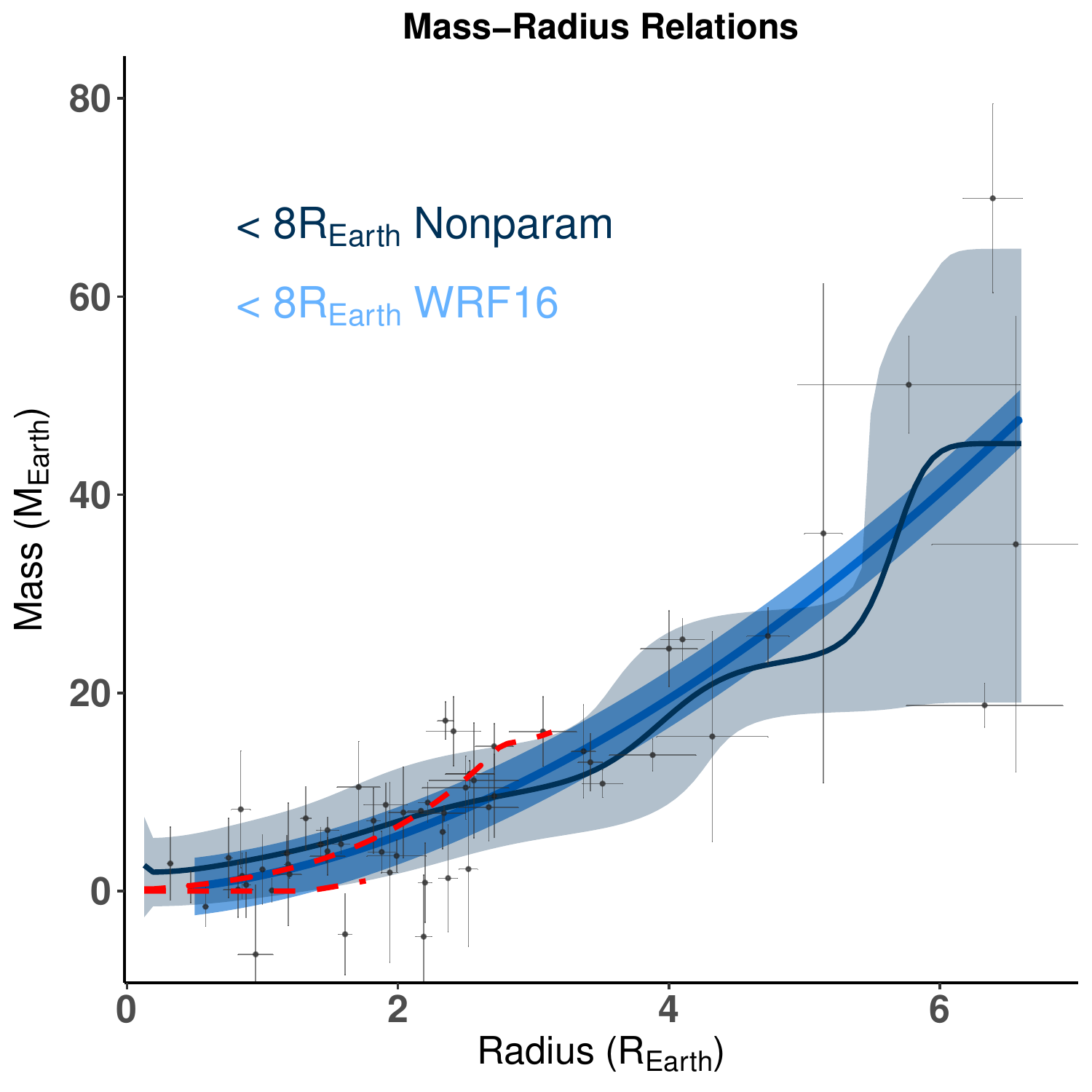}
   \caption{\small 
   Two M-R relations are plotted.
   The blue line is the mean power-law M-R relation estimated by WRF16{; the shaded region is the central 68\% of the predicted masses, showing the distribution of the mass predictions around the mean relation}.
   The dark blue line along with its {shaded} region is our nonparametric mean M-R relation and {the} 68\% prediction regions estimated using the Bernstein polynomial model.
Comparing the parametric M-R with the nonparametric M-R, the means of the M-R relations {noticeably differ in certain radius regimes}. 
The parametric model is optimistic with the assumption that mass and radius follow an increasing power-law relation even in the region where observations {are few and poorly constrain the relation}. The nonparametric M-R method is more adaptive to the observed data. 
The red lines {denote} physical restrictions for planet masses. Data points are displayed in the background with their measurement errors.
}
\label{fig:MRrelations}
\end{figure*}

As we look to apply our nonparametric method to a wider range of exoplanet masses and radii than considered by WRF16, it is nonetheless prudent to benchmark our results to their results.  

The parametric power-law M-R relation obtained by WRF16 is plotted in Figure \ref{fig:MRrelations}.
{Quantitatively, it is}
$m|r \sim \mathcal{N}(1.6r^{1.8}, 2.9)$ for $<8$\Rearth,
which assumes that the M-R relation follows a normal distribution.

In Figure \ref{fig:MRrelations}, we also plot the {mean and} 68\% prediction intervals of M-R relation estimated using the proposed nonparametric method (the grey region){.  Based on the 10-fold cross-validation method, 
the values for $\hat d$ and $\hat d'$ are chosen to be 
$50$ and $20$, respectively.  Physically motivated boundaries for any M-R relations are denoted by the red lines}. 
Those boundaries are calculated based on the constraint that 
$0 < M_i \leq M_{i,\text{pureFe}}$, $i = 1, \dots, n$,
where $M_{i,\text{pureFe}}$ is calculated using 
\citet{For07a, For07b}'s 
rock-iron internal structure models,
\begin{equation*}
	\log(M_{i,\text{pureFe}}) 
	=
	\frac{-b \pm \sqrt{b^2 - 4a(c- R_i)}}{2a},
\end{equation*}
with $a = 0.0975$, $b = 0.4938$, and $c = 0.7932$.

Here we shall digress from the discussion to introduce three different uncertainty regions which we will mention in the article: Bayesian credible intervals, prediction intervals and bootstrap confidence intervals. 
The Bayesian credible intervals are obtained using a Bayesian method. 
WRF16 obtains their Bayesian credible intervals from sample draws of the posterior distributions. 
For the nonparametric model used in this article, the uncertainty regions {are} called the prediction intervals {because} they are obtained from the prediction distribution of either radius given mass or mass given radius.
To be more explicit,
{the} prediction intervals are obtained from the distribution $f(m|r, \hat{\boldsymbol w}, \hat{d}, 
\hat{d}')$, where
$\hat{\boldsymbol w}$ is the maximum likelihood estimator (MLE) of $\boldsymbol w$. 
{The} prediction intervals do not account for variations on parameters $\hat{\boldsymbol w}$, $\hat{d}$ and $\hat{d'}$ themselves. 
To account for the variation of the parameters in the model, we use the bootstrap method. 
The bootstrap method is conducted 
by resampling (with replacement)
the observations $N$ times (here, we take $N = 100$). For each resampling, we obtain $\hat{d}$, $\hat{d'}$ and $\hat{\boldsymbol w}${.  Then} the bootstrap confidence intervals are obtained by {plugging in} $N$ sets of these estimated values into the M-R relation function one-by-one and {identifying the} 16\% and 84\% quantile{s of the mean relation over those $N$ realizations}.
Note that the Bayesian credible intervals, 
the prediction intervals and the bootstrap confidence intervals not only have different interpretations,
but also are derived from different models---the Bayesian credible intervals are obtained from WRF16's Bayesian hierarchical power-law model,
the other two intervals are obtained from our nonparametric model.
Due to these differences,
we will not compare those intervals.
{Instead}, 
we focus on comparing the mean M-R relations from each method.

\begin{figure}[!h]
\centering
   \includegraphics[width=8cm]{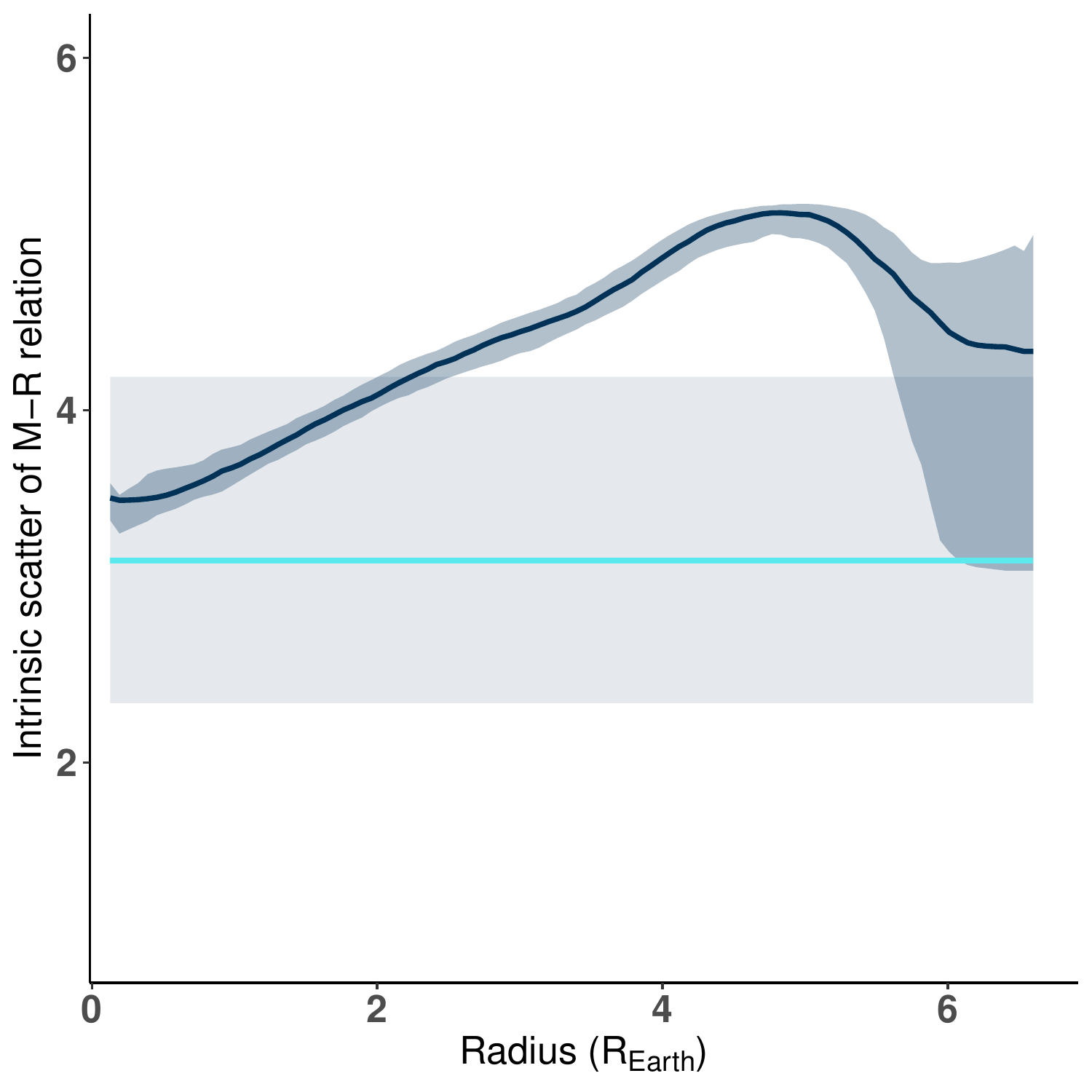}
   \caption{\small The intrinsic scatter is the estimated standard deviation (s.d.) for $f(m|r)$. The parametric model {(cyan)} is under-fitting the M-R relation, thus has a smaller {estimation for the s.d.\ compared} to the nonparametric model {(dark blue)}. The nonparametric model suggests that the {intrinsic scatter} is not a constant along with radii. In this figure,
the dark blue line plots {the mean s.d.\ across all bootstrapped realizations} estimated by the nonparametric method, {and the corresponding shaded region denotes} the bootstrap 16\% and 84\% confidence intervals. The {cyan line denotes the posterior median of the intrinsic scatter term in the parametric model; the associated shaded area denote the 16\% and 84\% quantiles of the posterior samples for that parameter}.}
   \label{fig:sdofMassgivenRadius}
\end{figure}

Now, let's go back to the discussion on Figure \ref{fig:MRrelations}. At the first glimpse, a distinct feature of this plot from the nonparametric M-R relation is that the uncertainty region {is} quite large {for $R>5$\Rearth compared} to the region obtained from the parametric M-R relation.
This is {because} there are only a few observations 
in that region,
and the measurement error for each observation is relatively large.
One may {worry that} the estimated mean M-R relation may {be misleading in this context}.
However, we view this as a strength of using the nonparametric model for at least two reasons: {first, there} is no concrete theory to support that the {population-level} relationship between mass and radius follows a power-law{, and so} the parametric M-R relation {may be} too optimistic with its precision in regions with little data. 
Second, 
from Figure \ref{fig:sdofMassgivenRadius},
the {constant variance assumed by the power-law model,
even in the $> 5$\Rearth regime},
may not reasonable.
The variance estimated using {our} nonparametric model {behaves more as expected: }the variance is smaller in the $<5$\Rearth region where the observations are abundant,
and it becomes larger in the $>5$\Rearth region.

In Figure \ref{fig:sdofMassgivenRadius}, we also plot the uncertainty regions for the estimated intrinsic {scatter from the two models, i.e.\ the 16\% and 84\% Bayesian} credible intervals using the parametric model, and 16\% and 84\% bootstrap confidence intervals using the nonparametric model.
Again, we would like to point out that the intrinsic scatter estimated by the nonparametric model is clearly not a constant.
In fact, WRF16 also believed that the intrinsic scatter may change as
planets increase the size.
However, {they only modeled the intrinsic scatter as a linear function of radius, and this model was not flexible enough to capture the true behavior:}
they found that the posterior distribution of the slope of that linear function {included zero, and thus} concluded that the intrinsic scatter was a constant with radius.

The normal assumption in the parametric model is also disfavored by observations.
From Figure \ref{fig:MRrelations}, the prediction intervals obtained from the nonparametric model suggests that
the conditional distribution of mass given radius is not {always} normally distributed as {there are regions where} those intervals are not symmetric around the mean.
When WRF16 checked the normal assumption after fitting their
model, they also found {evidence that} this assumption does not hold.

\section{The Joint Exoplanet Mass-Radius Distribution}
\label{sec:jointresult}

\begin{figure*}
\centering
   \includegraphics[width=13cm]{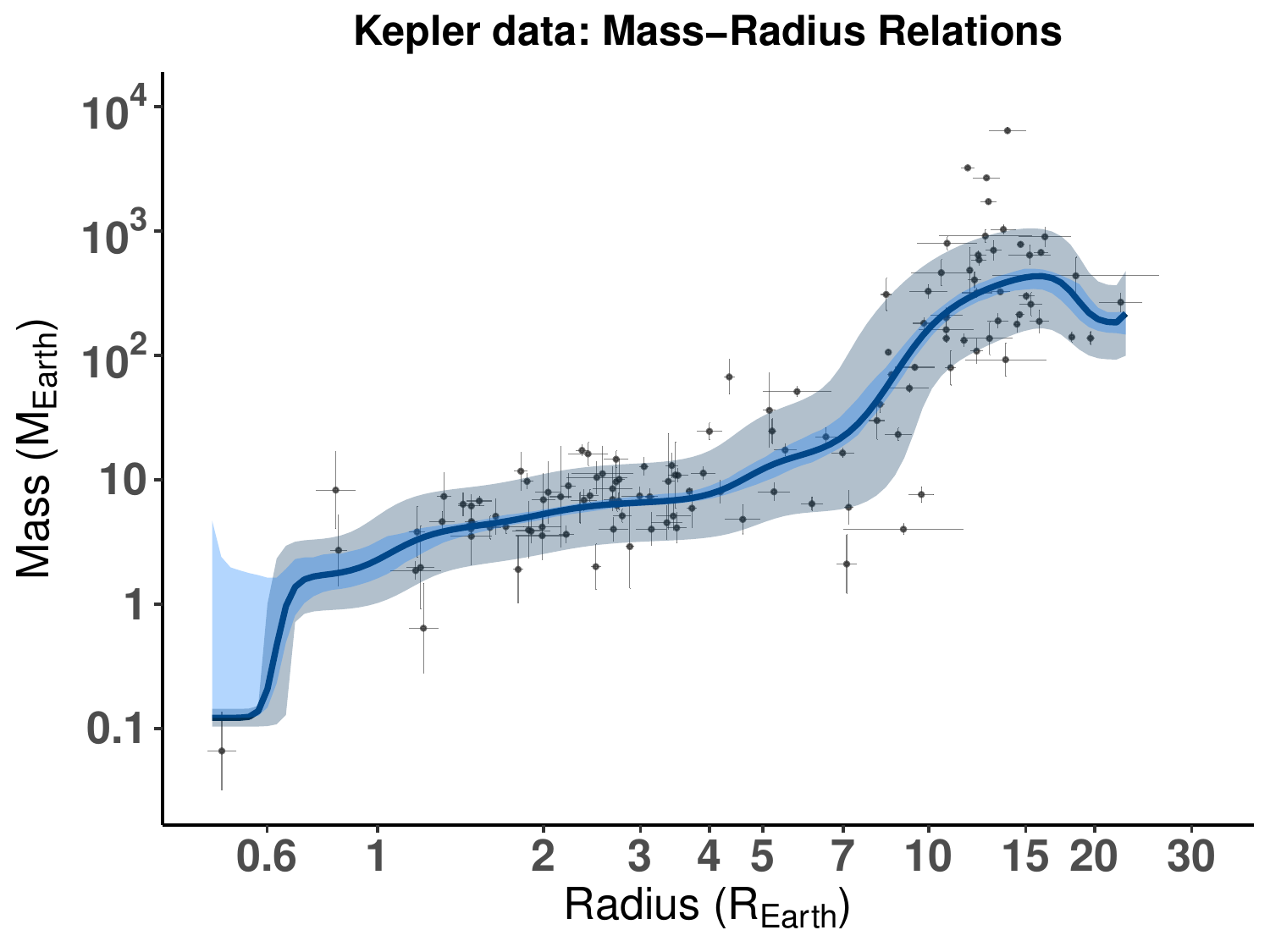}
   \caption{\small Nonparametric M-R relation of \Kep data.
   {With a more flexible model, we see that the transitions between super-Earths, Neptunes, and Jupiters are not sharply defined, yet there appears to be three contiguous regions between which the M-R relation changes: $0.8$\Rearth $\leq r \lesssim5$\Rearth, $5$\Rearth$ \lesssim r \lesssim11$\Rearth and $\gtrsim11$\Rearth}. The dark blue curve is the mean M-R relation and the {darker shaded} area is the uncertainty region between the 16\% and 84\% predictive intervals. The light blue band is the the uncertainty region between the bootstrap 16\% and 84\% confidence intervals {and shows large uncertainty in the region where data is absent (radius $<1$ \Rearth)}. Data points are displayed in the background with their measurement errors.}
    \label{fig:keplerMRrelation}
\end{figure*}

In this section, we repeat our nonparametric M-R relation analysis using all robustly measured
\Kep observations provided in the NASA {Exoplanet Archive, as described in \S \ref{sec:data}}. 
Since we consider a larger range of planet masses and radii,
we will display the M-R relation in log scale{ and} perform our analysis on $\log m$ and $\log r$,
rather than $m$ and $r$.
Note that the M-R relation in the {original} scale of masses and radii can be obtained by applying the Jacobian method
to the joint distribution of {$\log m$ and $\log r$}.

Similar to {Eqns \ref{eq:bernMod-A}--\ref{eq:bernMod-D}}, 
the joint distribution for {$\log m$ and $\log r$} is,
\begin{equation}
\label{eq:bernModLogScale}
\begin{split}
	f( & \log m, \log r| \boldsymbol w, d, d') 
	= \\
	& \sum_{k=1}^{d}\sum_{l=1}^{d'} w_{kl} 
	   \frac{\beta_{kd}
	   (\frac{\log m - \log\underline{M}}{\log\overline{M} - \log\underline{M}})
	   \beta_{ld'}(\frac{\log r - \log\underline{R}}{\log\overline{R} - \log\underline{R}})}
	   {(\log\overline{M} - \log\underline{M})
	   (\log\overline{R} - \log\underline{R})}.
\end{split}		
\end{equation} 
{Inferring the parameters $w, d, d'$ for} this model is similar {the inference performed for} the previous model. 

\subsection{M-R relation of the full \Kep dataset}

The M-R relation is the conditional distribution of {log-scaled} mass given {log-scaled} radius{, derived from Eqn \ref{eqn:cond}:} 
\begin{equation}\label{eq:condlogmlogr}
	f(\log m| \log r, \boldsymbol w, d, d') =
    \frac{f(\log m, \log r|\boldsymbol w, d, d')}{f(\log r| \boldsymbol w, d, d')},
\end{equation}
with $f(\log m, \log r|\boldsymbol w, d, d')$ is in
(\ref{eq:bernModLogScale}) and 
\begin{equation*}
f(\log r| \boldsymbol w, d, d')
= 
\sum_{k=1}^{d}\sum_{l=1}^{d'} w_{kl} 
	  \frac{\beta_{ld}(\frac{\log r - \log\underline{R}}{\log\overline{R} - \log\underline{R}})}{\log\overline{R} - \log\underline{R}}.
\end{equation*}

We use 10-fold cross-validation {(see \S \ref{sec:modinfer}) to select 
$d$ (corresponding to mass) and $d'$ (corresponding to radius), and find that $\hat d = 55$ and $\hat d' = 50$}.
After we obtain $\hat d$ and $\hat d'$,  we then {plug in} their values to estimate $\boldsymbol w$ using the MLE method. 
{Because} the values of $\hat d$ and $\hat d'$ may change {under different realizations of the random number generator, we repeat the 10-fold cross-validation method five times to assess its stability, each time using a different random seed.}
We found that four out of five times,
the cross-validation gives the same choices for 
$d$ and $d'${ with the dissenting realization suggesting} a slightly larger number for $\hat d$.
Thus we decide to use $\hat d = 55$ and $\hat d' = 50$ to estimate the M-R relation.
After we obtain $\hat d$, $\hat{d}'$ and $\hat{\boldsymbol w}$,
we derive the prediction distribution of the M-R relation{ and employ the bootstrap method described in \S \ref{sec:bench}} 
to obtain the bootstrap confidence intervals for the mean.

\begin{figure*}
\centering
    \includegraphics[width=8.5cm]{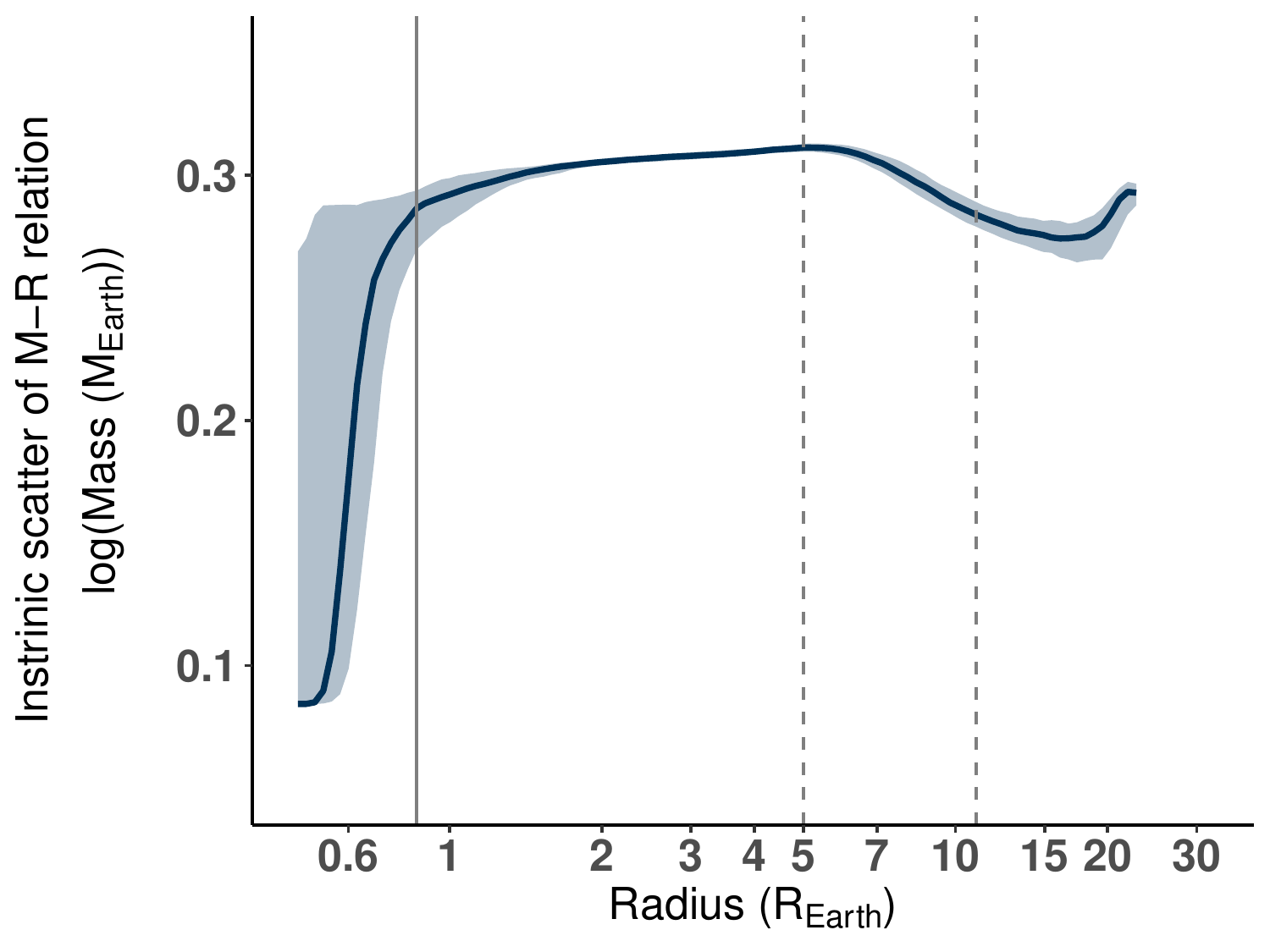}
    \includegraphics[width=8.5cm]{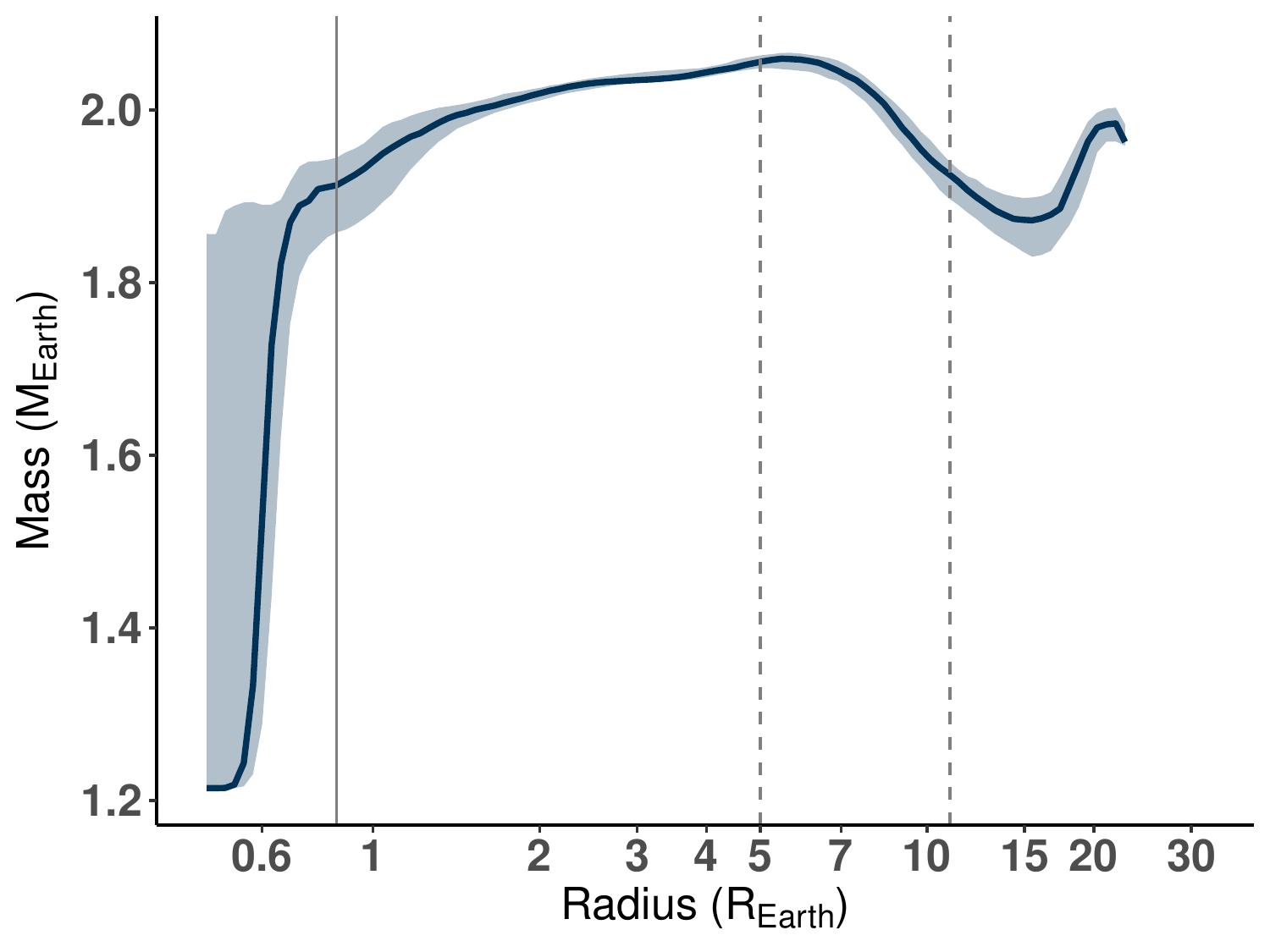}
   \caption{\small {The intrinsic scatter of the M-R relation, in log scale (left) and linear mass scale (right).  Mathematically,} the intrinsic scatter is the estimated {standard deviation of} $f(\log m|\log r)$. 
   We {note the behavior of the intrinsic scatter in the three regions visually identified in Fig \ref{fig:keplerMRrelation}: for $0.8$\Rearth $\leq r \lesssim5$\Rearth the intrinsic scatter is increasing; for $5$\Rearth$ \lesssim r \lesssim11$\Rearth it starts to decline; and for $\gtrsim11$\Rearth it is nonlinear.}
In the figure, the dark blue curve is the estimated {standard deviation of the M-R relation and the shaded} area is the uncertainty region between the 16\% and 84\% bootstrap confidence intervals.}
\label{fig:keplersd}
\end{figure*}

We plot the mean along with the prediction intervals and the bootstrap confidence intervals in Figure \ref{fig:keplerMRrelation}. 
In the plot,
the gray region represents the 16\% and 84\% prediction intervals and the blue region represents the 16\% and 84\% bootstrap confidence intervals for the mean. 
The black line represents the estimated M-R relation.
From the plot, we {see} that 
the M-R relation {below $0.8 $\Rearth is unexpectedly similar to a step function}.
This is because there is only one {isolated} data point in that region{, which} the model tries to connect {with the nearby} points {via} smooth polynomials.
{Fortunately, we see that} the bootstrap confidence intervals in this region {are} very wide{, indicating that the M-R relation is not well understood in this region, and that more data is needed}. 

The M-R relations can be roughly split into three parts{: $0.8$\Rearth $\leq r \lesssim5$\Rearth, $5$\Rearth$ \lesssim r \lesssim11$\Rearth and $\gtrsim11$\Rearth}.
In the region between $0.8$\Rearth and $5$\Rearth, the bootstrap confidence intervals narrow as data points become abundant for larger radii, which suggests the M-R relation is more accurate.
The almost linear relation suggests that the power-law relation may not be a misleading assumption {in this radius regime}.
{Transitioning} from the $5$\Rearth region to the $11$\Rearth region, 
the number of observations decreases and we observe both the prediction intervals and bootstrap confidence intervals become slightly wider. 
In the region $>11$\Rearth the M-R relation is more flat or even decreases as radius becomes larger. 
These findings are {consistent with well-known features of the M-R relation (see discussion in \S \ref{sec:wellunderstood}).}

{Figure \ref{fig:keplersd} shows} how the intrinsic scatter changes {as a function of} radius{; because the intrinsic scatter is defined as the standard deviation of $f(\log m|\log r)$ (see Eqn \ref{eq:condlogmlogr}), the intrinsic scatter is technically in units of $\log m$.  We plot this on the left and original scale on the right for a more intuitive comparison}.
{Starting at the smallest radii, we see that the bootstrap confidence intervals are quite wide.  This reflects the fact that there is only one data point below $\sim 0.8$\Rearth, and so the intrinsic scatter is quite uncertain in this regime.}
In the region between $0.8 $\Rearth and $\lesssim 5$\Rearth,
the intrinsic scatter is an increasing function with radius. The increasing pattern becomes more obvious in the right plot. 
This finding is consistent with the result shown in Figure \ref{fig:sdofMassgivenRadius},
even though two results are based on different datasets. 
With more data points {to constrain this region, we see that} the size of the intrinsic scatter in Figure \ref{fig:keplersd} is smaller than it in 
Figure \ref{fig:sdofMassgivenRadius}.
In the $>5$\Rearth region, the intrinsic scatter behaves nonlinearly. It decreases first {then increases at radii dominated by inflated Jupiters.  As discussed in \S \ref{sec:wellunderstood}, the M-R relation is difficult to interpret in the Jupiter regime because degeneracy pressure creates a vertical line in this space that is not well captured by the conditional.  See the R-M relation in Fig \ref{fig:keplerRMrelation} for a clearer picture of the model's behavior in the Jupiter regime.}

\subsection{{Mass predictions for given radii}}

\begin{figure*}
\centering
    \includegraphics[width=14cm]{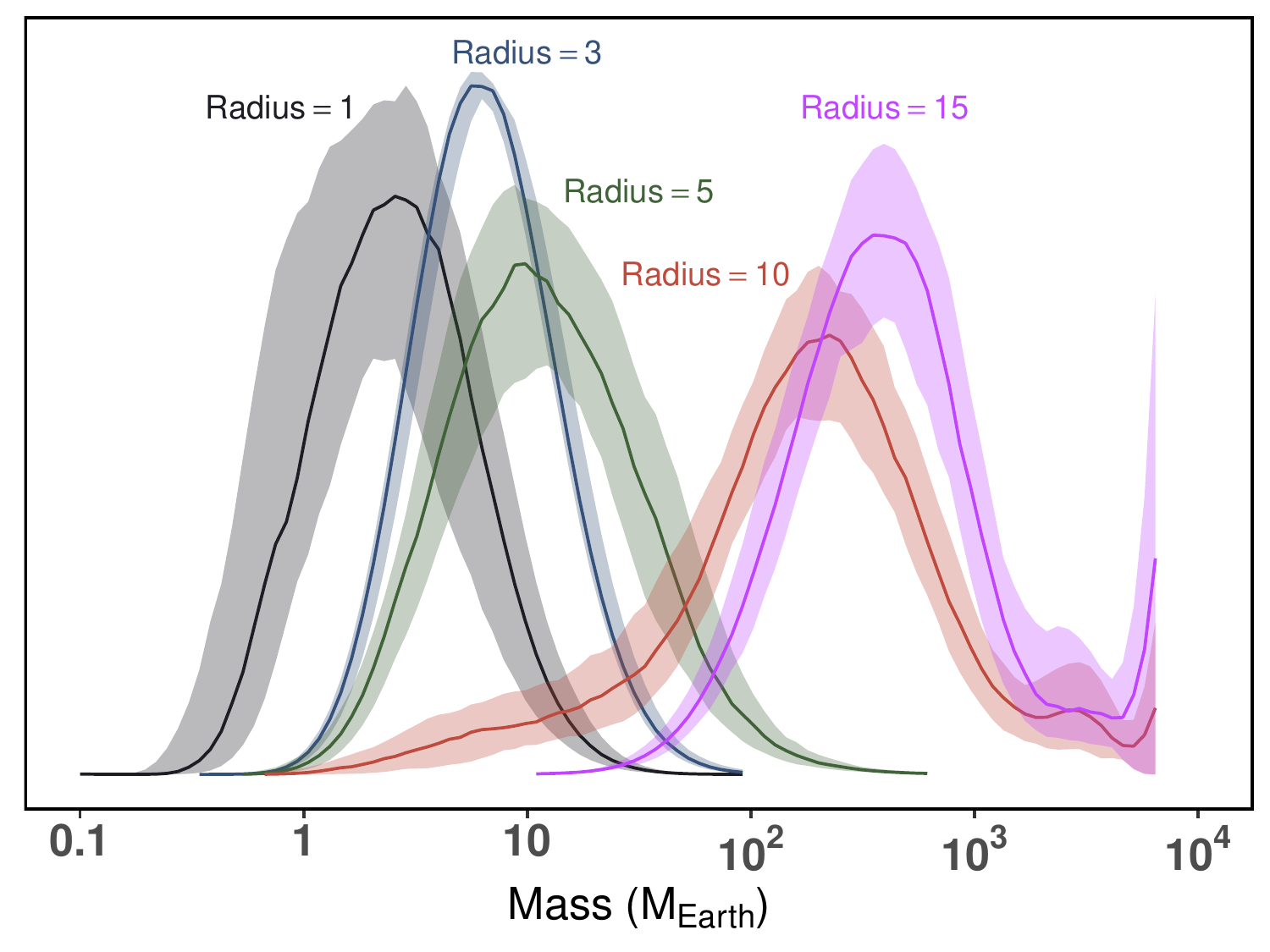}
   \caption{\small The conditional distributions for mass given radius{, i.e.\ the mass predictions that our model produces for planets at 1\Rearth, 3\Rearth, 5\Rearth, 10\Rearth, and 15\Rearth.  In the plotted log scale, these predictive mass distributions are roughly Gaussian, but in linear scale are skewed} to the right. The larger radius is, the more {skewed the distribution}. 
{For each curve, the shaded region corresponds to the uncertainty in the distribution, i.e.\ the} 16\% and 84\% bootstrap confidence intervals.}
\label{fig:keplerconddens}
\end{figure*}

In Figure \ref{fig:keplerconddens}, 
we plot five conditional densities for mass given radius at different radius values: $1$\Rearth, $3$\Rearth, $5$\Rearth, $10$\Rearth, {and} $15$\Rearth.
{These curves represent the mass predictions for planets at those radii, i.e.\ the probability of a planet of that size having a certain mass.  The shaded regions represent the uncertainty in that distribution and correspond to} the 68\% bootstrap confidence regions.
{While these distributions look roughly Gaussian in log scale, when transformed to a linear scale} all the densities are skew{ed} to the right. 
The two densities within the region $>5$\Rearth are more skewed than the rest. 
Thus no matter the the size of the planets, the intrinsic scatter is definitely not Gaussian {in linear scale, as was assumed by WRF16.  The lognormal intrinsic scatter assumed by \citet{Chen17} is more appropriate for mass given radius, although it fails to capture the low-mass tail corresponding to super-puffy planets at $7-10$ \Rearth and $<30$ \Mearth.}

\subsection{The radius{-mass (R-M)} relation}

\begin{figure}[!h]
\centering
   \includegraphics[width=8.5cm]{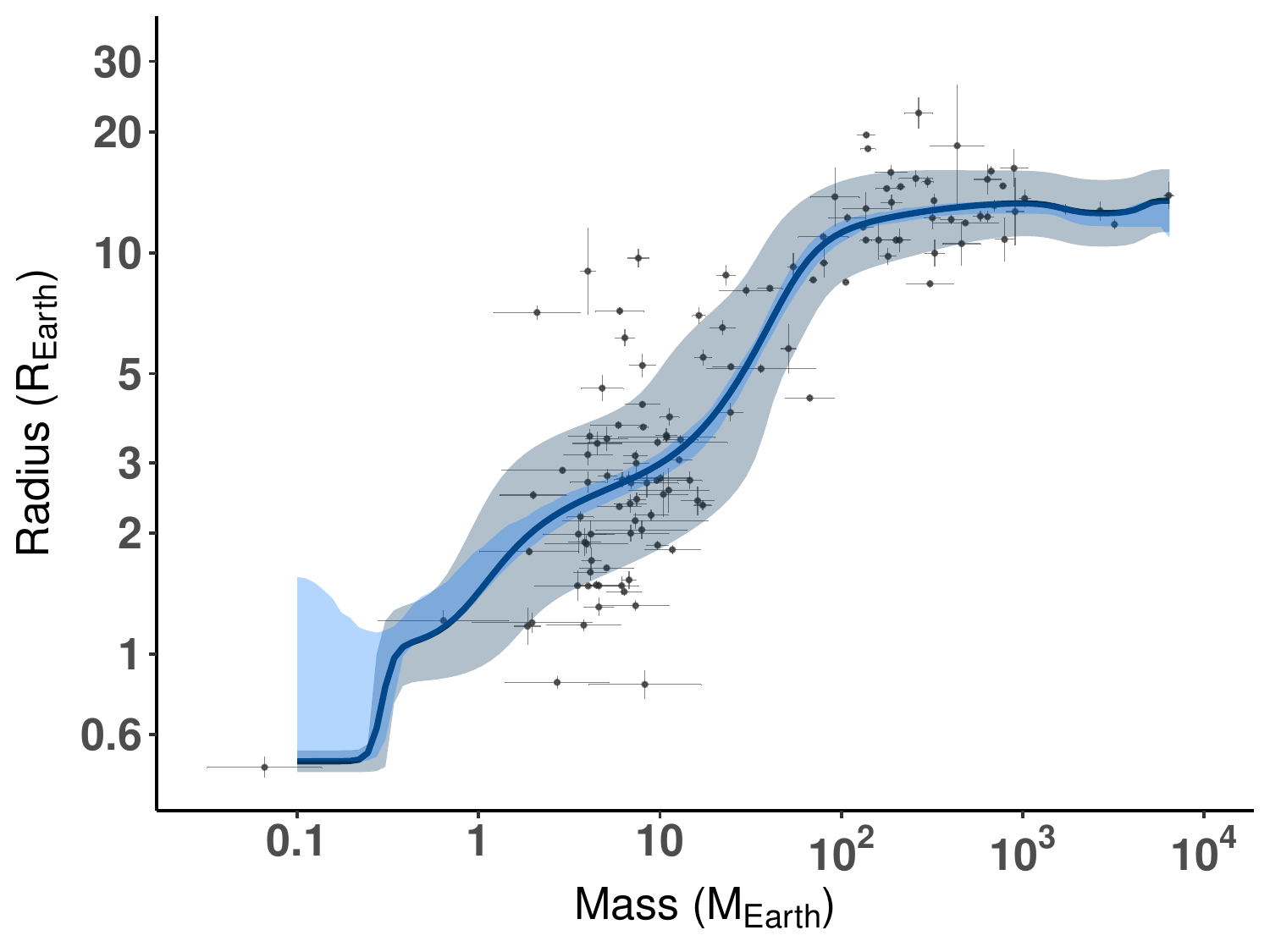}
\caption{\small The nonparametric estimation of {the} R-M relation{; note that this is not simply the inverse} of the M-R relation. The dark blue line is the mean R-M relation and the gray area is the 16\% and 84\% prediction intervals. The light blue area is the uncertainty region between 16\% and 84\% bootstrap confidence intervals.}
\label{fig:keplerRMrelation}
\end{figure}

As we are modeling the joint distribution of mass and radius (in log scale), the radius{-}mass relation (R-M relation) can also be derived easily. From Eqn \ref{eqn:cond-rm}, 
the conditional distribution for $\log r$ given $\log m$ is
\begin{equation}
	f(\log r| \log m, \boldsymbol w, d, d') =
    \frac{f(\log m, \log r|\boldsymbol w, d, d')}{f(\log m| \boldsymbol w, d, d')},
\end{equation}
with $f(\log m, \log r|\boldsymbol w, d, d')$ in
(\ref{eq:bernModLogScale}) and 
\begin{equation*}
f(\log m| \boldsymbol w, d, d')
= 
\sum_{k=1}^{d}\sum_{l=1}^{d'} w_{kl} 
	  \frac{\beta_{ld}(\frac{\log m - \log\underline{M}}
      {\log\overline{M} - \log\underline{M}})}
      {\log\overline{M} - \log\underline{M}}.
\end{equation*}
After we {plug in} $\hat d$, $\hat d'$ and $\hat{\boldsymbol w}$,
we plot this relation along with the prediction intervals and bootstrap confidence intervals in Figure \ref{fig:keplerRMrelation}.
{Note} that the R-M relation is not a simple flip of the M-R relation {around the 1:1 line; this is due to the fact that the M-R relation is actually a distribution that is asymmetric around a mean relation that is not a straight line.}

\begin{figure}[!h]
   \includegraphics[width=9cm]{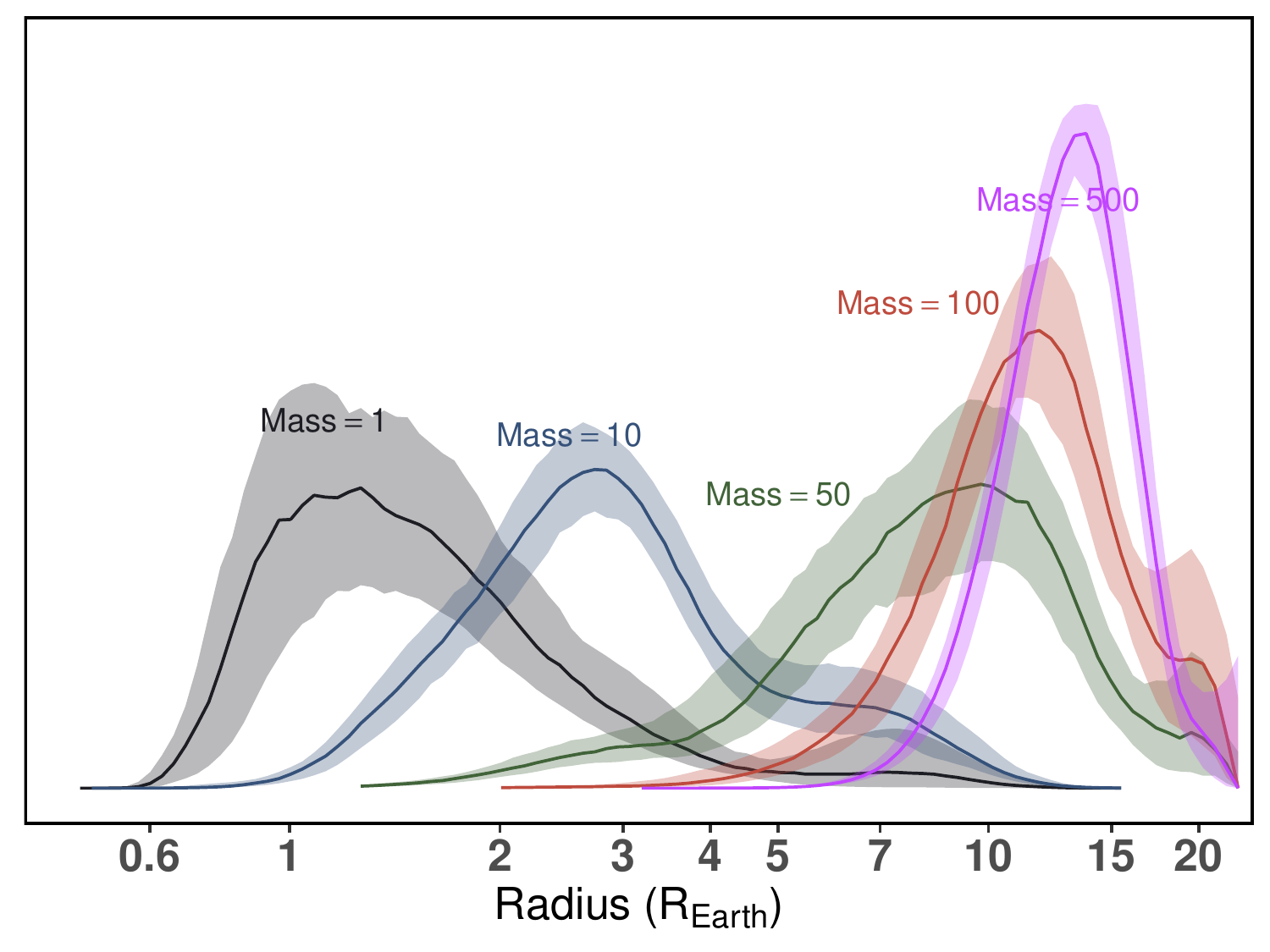}
   \caption{\small The conditional distribution{s} for radii given masses {at} 1, 10, 50, 100, 500 $M_\oplus$. {As in Figure \ref{fig:keplerconddens}, the shaded region corresponds to the uncertainty in the predictive distributions, i.e. the} 16\% and 84\% bootstrap confidence intervals.  {These predictive distributions are more skewed than those for mass given radius.}}
    \label{fig:keplerCondRM}
\end{figure}

The R-M relation is {highly uncertain for} $<0.8M_\oplus$ {due to the single isolated data point} in that region{; this is reflected in} the wide bootstrap confidence intervals.
The bootstrap confidence intervals narrow {as mass increases, reflecting higher certainty in the mean relation given the data}. 
{This mean}
 R-M relation is an increasing function {up to 100 \Mearth; above that}
 the R-M relation becomes flatter {due to degeneracy pressure}.

We also plot the 68\% bootstrap confidence intervals for each of the conditional distribution{s for} 1, 10, 50, 100 and 500 {\Mearth.}
Almost all the densities are skew{ed} to the right{, especially when the displayed log scale is transformed to linear scale}. 
In the $0.8M_\oplus \leq m \leq  10^2 M_\oplus$ region,
the three densities look quite different{, indicating rapid change in the radius distribution from 1 to 50 \Mearth}.

\subsection{Predicting K2 planets}

\begin{figure}
\centering
    \includegraphics[width=8.5cm]{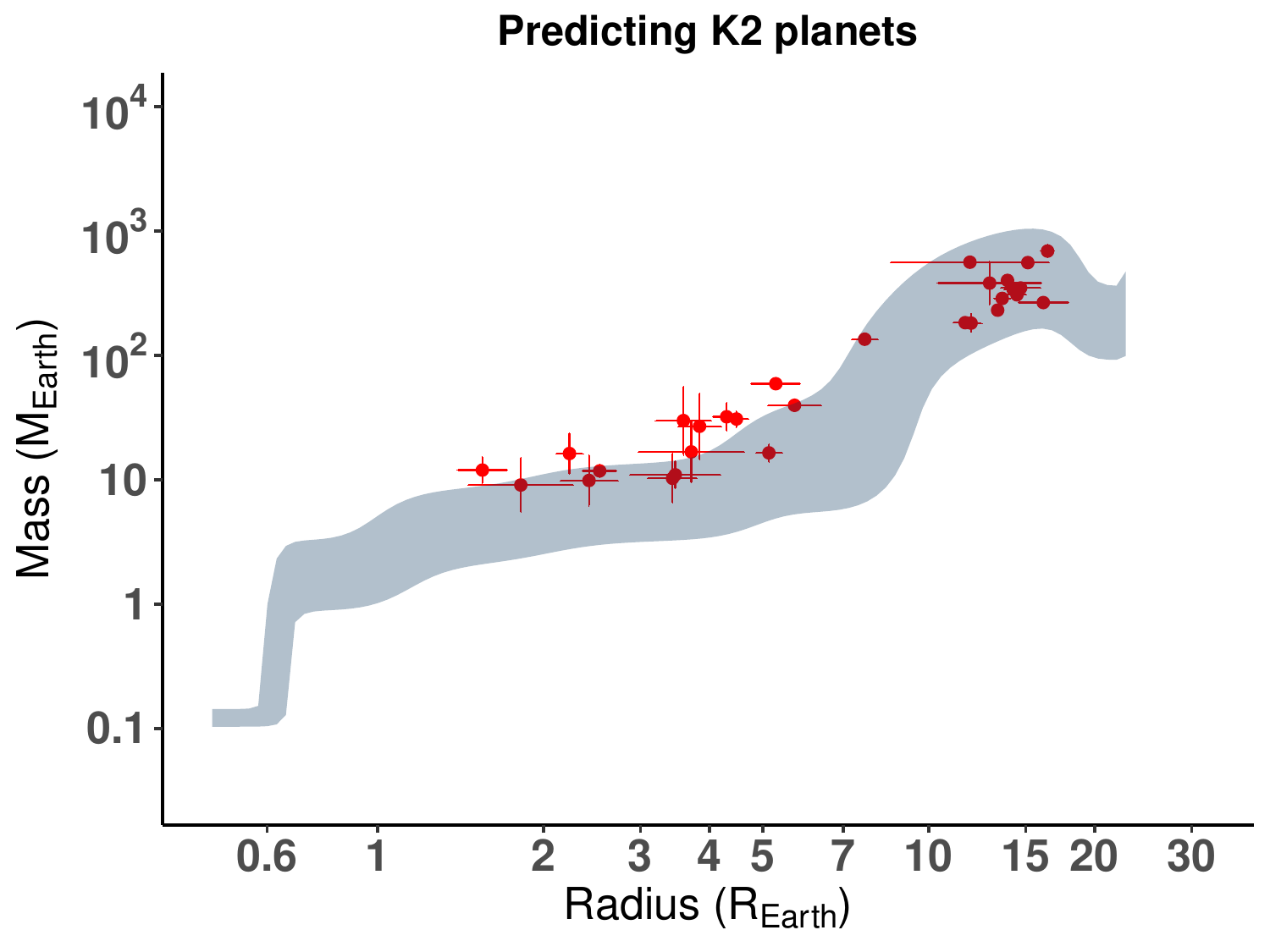}
    \caption{\small {The masses of} K2 planets with radii $<10 $\Rearth are biased. The shaded area is the {same as that shown in Figure \ref{fig:keplerMRrelation}; it is} the uncertainty region between 16\% and 84\% prediction intervals of {the} M-R relation estimated using \Kep data. K2 planets{' masses and radii} are plotted {in red} with their measurement errors.}
    \label{fig:predictK2}
\end{figure}

In this section, we apply the estimated M-R relation to predict masses of K2 planets given their radius. 
The K2 planets data are obtained from the NASA Exoplanet Archive on June 5, 2017. 
In Figure {\ref{fig:predictK2}},
we first plot the prediction intervals {from} the M-R relation for all {radii},
then plot the K2 planets on top of the band.
{We} see that for many of the planets with $R<10 $\Rearth, the prediction intervals do not cover their masses. However,
for planets with $R>10 $\Rearth, the prediction intervals {do span the} masses for all the planets.
{This} suggests {that there is a significant} bias for K2 planets{.  One explanation for this is that the most massive planets reach a high mass detection significance threshold first, and so are published first.  \citet{Burt18} discusses this bias and its implication for population mass-radius analyses in detail.} 

\section{Implications for Exoplanet Compositions}
\label{sec:disc}

The nonparametric M-R relation that we have fit to the full \Kep dataset yields several astrophysical results.
{Below we} discuss them in detail\sout{s}.

\subsection{Assessing ``well-understood" M-R features}\label{sec:wellunderstood}

A crucial test of any method is how well it reproduces features that are well established by multiple studies using different techniques.  {To this end, we emphasize that our relation is not just the average line,
but also the predictive distribution around the line.  We also note that our goal was not to reproduce prior mass-radius relations, as} one of our motivations for applying a nonparametric method {was} to assess how well the features identified in more parametric models persist when the strong assumptions in those models are relaxed. 

{To start with, our nonparametric M-R relation reproduces the well-established result that} more massive planets $\lesssim 10^2$\Mearth are on average larger {(see Figure \ref{fig:keplerMRrelation})}. This result is not a surprise: more massive planets are expected to form earlier in the protoplanetary disk lifetime and thus should able to accrete more gas-rich material, which quickly increases the planets' radii.  
Indeed, all M-R relations in the literature (see \S \ref{sec:litM-R}) show that mass (or radius) is an increasing function of radius (or mass) in the sub-Jupiter regime. 
However, above about half the mass of Jupiter, degeneracy pressure begins to noticeably affect the radii of hydrogen-dominated bodies \citep{Zap69}, and the R--M relation should flatten out.
Our nonparametric approach readily reproduces this result (see Figure \ref{fig:keplerRMrelation}).

The electron degeneracy pressure would manifest as a
vertical line in the M-R relation of Figure \ref{fig:keplerMRrelation}; when taking the marginal of the joint mass-radius
distribution to produce the $f(m|r)$ (mass given radius) conditional distribution, vertical features like
this are collapsed, with their extent represented as the probability distribution at that radius.
Figure \ref{fig:keplerCondRM} displays these probability distributions, and shows that there is some probability of
more massive planets at 10 and 15$R_\oplus$. In fact, the difference between $f(m|r)$ and $f(r|m)$ in the
Jupiter regime is an argument for why a two-way M-R relation is necessary: there are some
features that are easier to capture one way than the other.

\subsection{Location of transition points}

Identifying transitions in the exoplanet mass-radius relation is important for our physical understanding of the types of planets which exist, based on their compositions.

{No sharp transitions are visible in Figure \ref{fig:keplerMRrelation}.  Nevertheless, there appears to be at least three segments to the M-R relation:
 $\lesssim 5$\Rearth, $\sim 5$ to $\sim 11$\Rearth, and $\gtrsim 11 $\Rearth radii, which roughly corresponds to sub-Neptunes, sub-Saturns, and Jupiters}.
 A comparison of Figure \ref{fig:keplerMRrelation}, the M-R relation, to Figure \ref{fig:keplerRMrelation}, the R--M relation, shows that there is uncertainty in the {sub-Neptune to sub-Saturn} transition point.  
 This is because we relax the assumption that the distribution around the average relation is symmetric and constant within a segment, as is assumed by WRF16 and \citet{Chen17} via both studies' use of a Gaussian distribution to characterize the intrinsic scatter term.  With this restriction lifted, the average mass given radius is no longer the inverse of the average radius given mass, as the conditionals shown in these two plots are produced by marginalizing over a joint mass and radius distribution that is asymmetric.  When the joint distribution is much wider in one dimension than it is in the other, as is the case around 4-7\Rearth, then the conditionals can look quite different.  A relatively large spread in radii for planets with 3-10\Mearth falls adjacent to a small spread in radii for planets around 30\Mearth; this is what causes ambiguity in identifying a transition from Neptunes to Saturns.

Upon visual inspection, it is also clear that there are not currently enough mass measurements of extrasolar planets $\leq 2$\Rearth to justify a super-Earth transition point.  This result is at odds with some previously published results on the M-R relation, 
for example  
\citet{Wei14}, \citet{Rog15}, \citet{Chen17} and \citet{Ful17}.
In below, we shall discuss each of their works one by one.

\citet{Wei14} fixed the transition point at 1.5$R_\oplus$ 
based on visual inspection of the density vs. radius plane. 
{When their dataset is plotted with the reported error bars, it becomes apparent that} there is little empirical
evidence for a transition at 1.5$R_\oplus$ {given the
size of the density or mass error bars for the smallest planets}. 
Indeed, one could draw a straight line through the mass and radius points and have it fit the data just as well as the chosen
relation with the transition. 

To our best knowledge, \citet{Rog15} provided the strongest evidence of a transition in the M,R plane among all {four of these studies}. 
{This paper} parameterized the transition as
the radius below which 100\% of planets must be rocky (that is, their masses are greater
than the minimum mass of a rocky planet at that size). {This study tested} both a step function for
the fraction of planets that are rocky and a more gradual transition; given
{the size of the error bars}, a step function was sufficient {to describe this dataset}. While useful in its physical interpretation, this
parametrization does not guarantee that there is a {visible} kink {or transition} in the M-R relation. Indeed, no
such kink is visible in their dataset, displayed in their Figure 1. This is because the
Rogers parameterization of the transition tests for the absence of planets in one region of
M,R parameter space (that of very low-mass planets in the  $<$ 1.5$R_\oplus$ radius
regime); if their absence does not sufficiently alter the average M-R relation, then it will
not produce a kink in the M-R plane. Considering the very large measurement
uncertainties in this very small planet regime, it is indeed the case that the absence of
these planets {doesn't} significantly shift the average M-R relation to produce a kink.

\citet{Ful17} looked only at the marginal radius distribution. There is no guarantee
that the bimodality they see in the radius distribution will produce a kink, or a transition, in the M-R plane. Indeed, the transition is likely gradual with a period dependence.
Additional follow-up to measure masses for a larger sample of these small planets is
needed before {a claim of} a Fulton-like gap or transition in the M-R plane {is warranted}.

{The tension between the existence of a super-Earth transition in \citet{Chen17} and our lack of one} arises from the fact that we do not include the minor Solar System bodies in our dataset.  It is not unreasonable to expect that the compositions of mainly refractory bodies in other planetary systems will be similar to those in ours.  However, including this data without inflating the uncertainties to match those we obtain for extrasolar systems leads to misleadingly tight constraints for $\sim 1$\Mearth planets.  Indeed, there is only one planet between 1 and 3\Mearth in {their} dataset, so most of the information about the ``Terran" transition point comes from extrapolating up from Solar System minor bodies and extrapolating down from Neptunes, rather from measurements at the transition point itself.  We choose a less Solar System-constrained approach to characterize the distribution of compositions in the 1-3\Rearth regime{, choosing instead to develop} a flexible method that can best let the extrasolar data speak for itself as more super-Earths are discovered and followed up by TESS.

\subsection{{Intrinsic dispersion}}

The intrinsic dispersion in mass given radius {slightly increases from 1 to 5 \Rearth} (see Figure \ref{fig:keplersd}), but varies much more in radius given mass {(see the width of the gray region in Figure \ref{fig:keplerRMrelation}}.  This indicates that planet compositions in the $3$-$10$\Mearth range exhibit more diversity than others, and that planets with $1$-$5$\Rearth fall into a relatively narrow range of masses.  The large spread in radius over a narrow mass regime could have a number of physical interpretations: perhaps these super-Earths/sub-Neptunes fall at a transition point where scaled-up terrestrial planet formation co-exists with scaled-down gaseous planet formation.  Perhaps planet evolution is particularly dramatic in this mass regime; photoevaporation has been shown to be important for such low-density low-mass planets \citep[e.g.][]{Lop16,Owen17}.  

In either case, it would be useful to search for multiple populations in this region of the M-R diagram, to test if the bimodal marginal radius distribution unearthed by \citet{Ful17} and verified by \citet{Hsu18} extends to the joint radius and mass space.  While there is not currently evidence for a bimodal mass and radius distribution, our approach of calculating the M-R and R--M relations by first estimating the joint distribution is capable of finding and quantifying such bimodalities. Before the joint distribution can be interpreted as an astrophysical result, however, both selection effects discussed in \S \ref{sec:seleffect} would need to be incorporated into the density estimation.  

\subsection{{Predictive distributions}}

Given the current dataset, a {lognormal} distribution is a reasonable approximation for {the conditional distribution of mass given radius, in} most radius regimes ({see Figure \ref{fig:keplerconddens}}).  On the other hand, there is significant skewness in the conditional distributions for log scaled radius given log scaled mass (see Figure \ref{fig:keplerCondRM}). This result demonstrates that adopting the more flexible model was warranted, and that different mass regimes have different composition distributions.  In particular, the super-puffy planets found by TTV analyses are visible as the skewed tail in the 10\Mearth conditional distribution.

\subsection{Implications for TESS}

As seen in Figure \ref{fig:keplerconddens}, mass predictions are the best constrained for planets at 3\Rearth: the conditional distribution for mass at 3\Rearth is the most robust to the bootstrapping scheme we used to measure the uncertainty in these distributions.  As a result, RV follow-up of TESS will yield the most scientific return for 1-2\Rearth planets and above $\sim 4$\Rearth where there are intrinsically fewer planets.

While few \Kep target stars were sufficiently bright to enable follow-up, a larger proportion of K2 planet-hosting stars are.  One would hope that these brighter stars would enable more mass detections of smaller planets, yet we see in Figure \ref{fig:predictK2} that there is significant bias in the K2 planets planetary masses, especially for the planets with $\leq 7$\Rearth radii.  This serves as a word of warning for those who aim to perform mass-radius analyses with TESS data: to build robust M-R relations, {all mass measurements, even upper limits, must be published and incorporated into the analysis (see \citet{Burt18} for an in-depth study of this effect).  In addition,} the second type of selection effect described in \S \ref{sec:data} must be quantified and published as part of major radial velocity follow-up programs.

\section{Conclusion}
\label{sec:conc}

In this paper we provide a nonparametric approach,
using the Bernstein polynomial model,
to estimate the M-R relation.
We applied this approach to analyze two datasets.

The first dataset is from WRF16{, which we use to benchmark our results}. 
First, we found that the M-R relations estimated using the 
parametric model and {the} nonparametric model are similar for 
$<5$\Rearth planets.
{However, the two relations differ for} $>5$\Rearth{, in that the} prediction intervals of the M-R relation obtained
from the nonparametric model are wider from a model that 
assumed a power-law M-R relation. 
We also found that the conditional distribution 
of mass given radius is not {always} normally distributed.
Last, we found that the intrinsic scatter is not a constant function with radius{, increasing weakly up to 5\Rearth}. 

We applied the Bernstein polynomial model to study the joint exoplanet mass and radius distribution 
using all \Kep planets with measured masses.
We found that there are at least three distinct M-R relations for planets with radii $\lesssim 5$\Rearth, $5$\Rearth $< r <11$\Rearth and $>11$\Rearth.
We also studied the relationship between the
intrinsic scatter and the radius.
We also found that the intrinsic scatter is \textbf{a weakly} increasing function with radius { up to $\sim 5$\Rearth and becomes nonlinear beyond that.}
Furthermore, we found that the conditional distribution of mass given radius is {reasonably approximated by a lognormal} distribution, which skews to the right in $m$, but {is} nearly symmetric in $\log m$. 
Last, we studied the R--M relation
and applied the estimated M-R relation to predict K2 planets. We found that the bias for K2 planet{ masses},
especially for {those with $<10$\Rearth,} is large.

A major contribution {of} this study is that our method provides a tool to explore a wider range of fits by incorporating {heteroscedastic} measurement errors. Our method provides a new direction for {studying} the M-R relation when more mass and radius measurements are obtained {with future missions like TESS and PLATO}.
{To enable these future studies, we have placed the input datasets and the $\textsf{R}$ code which we use to fit the weights, find the optimal number of degrees, and plot the figures in this paper, at the following website}:
\href{https://github.com/Bo-Ning/Predicting-exoplanet-mass-and-radius-relationship}{https://github.com/Bo-Ning/Predicting-exoplanet-mass-and-radius-relationship}.

In the future,
the model itself can be extended by 
adding 
incident flux as a third variable for a better understanding of the M-R relation
(see \citet{Tho17} and \citet{Ses18} {for examples of flux-dependent M-R analysis in the Jupiter regime}).
{It is likely that a mass-radius-flux} (M-R-F) relation
would better describe planets in all mass regimes, given the importance of photoevaporation for
low-mass planets.  Given the rich and
growing literature on the subject, {this is an area where significant progress can be made, with the right tools.  As we have demonstrated in this paper, a nonparametric approach provides such a tool: it can yield numerous astrophysical insights and model the data with minimal assumptions, making it a promising option for future mass-radius analyses.}

\acknowledgments

The first draft was completed when the first author was a Ph.D. student in Department of Statistics at North Carolina State University, Raleigh, North Carolina, U.S.A..

The authors thank the Statistical and Applied Mathematical Sciences Institute (SAMSI) for bringing the authors together and providing space and funding for continued collaborations. The authors also thank Tom Loredo and Eric Ford for their valuable suggestions on this work while it was in development at SAMSI, {Eric Ford} for providing detailed revisions on the first version of the manuscript{, Shubham Kanodia and Gudmundur Stefansson for their work in translating the R code into Python, and Matthias Yang He and Eric Ford for their work in testing the number of effective nonzero weights while translating the code into Julia.}
Furthermore, the authors thank two referees who provided valuable suggestions to improve the paper.
This material was based upon work partially supported by the National Science Foundation under Grant DMS-1127914 to the Statistical and Applied Mathematical Sciences Institute.  AW acknowledges support from the National Science Foundation Astronomy \& Astrophysics Postdoctoral Fellowship program under Award No. 1501440.  
The Center for Exoplanets and Habitable Worlds is supported by the Pennsylvania State University, the Eberly College of Science, and the Pennsylvania Space Grant Consortium.  
This research has made use of the NASA Exoplanet Archive, which is operated by the California Institute of Technology, under contract with the National Aeronautics and Space Administration under the Exoplanet Exploration Program. This paper includes data collected by the \Kep mission. Funding for the \Kep mission is provided by the NASA Science Mission directorate.

\vspace{5mm}
\facilities{NASA Exoplanet Archive}

\software{\textsf{R} and python}

\appendix

\section{Bernstein Polynomials and their Properties}
\label{sec:BPprop}

\begin{figure}[!h]
\centering
   \includegraphics{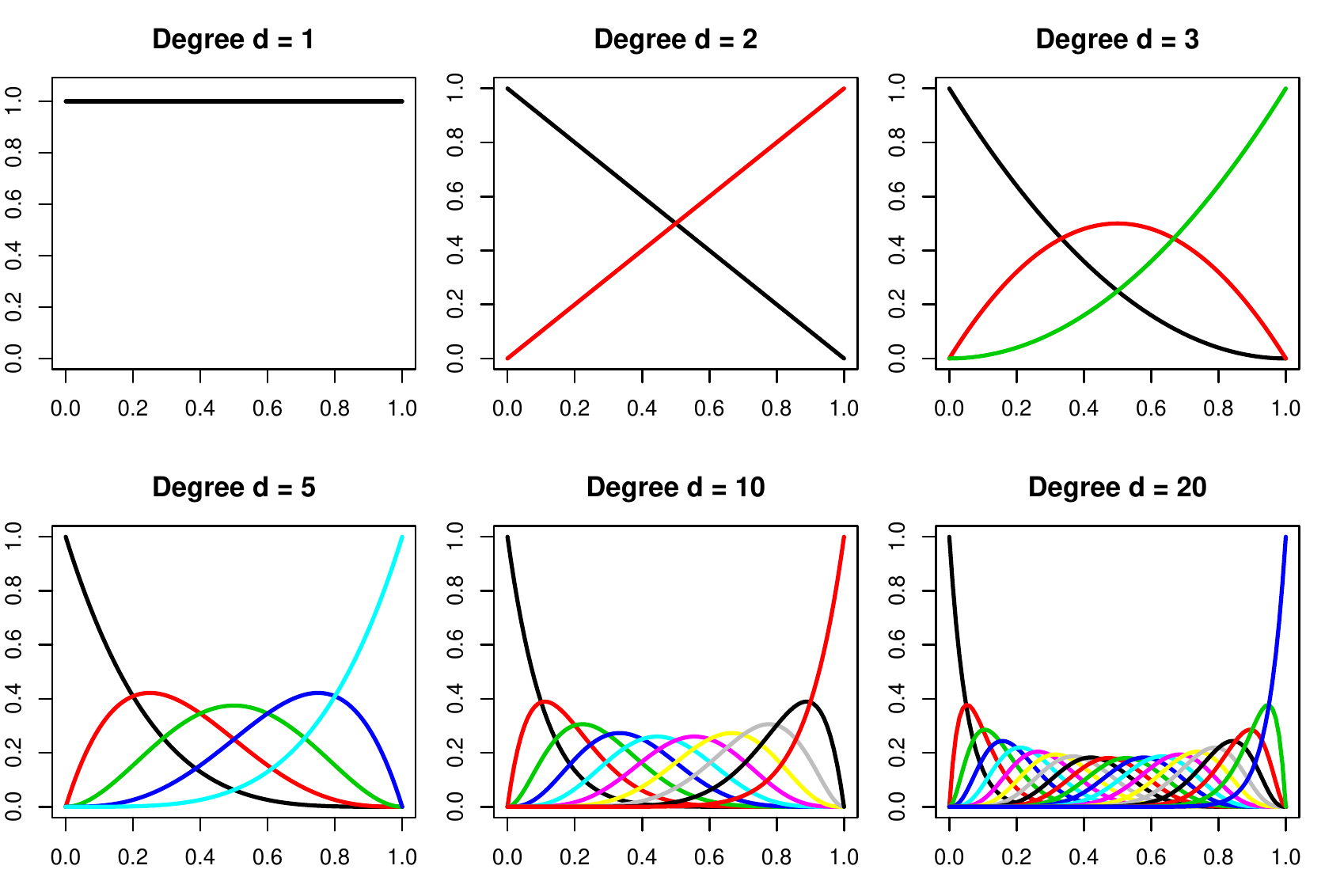} 
   \caption{\small Plot of Bernstein polynomial basic functions{, as a function of degree (i.e.\ the number of terms in the polynomial series); compare to Eqn \ref{eq:beta}.  Note that the independent variable, i.e.\ the x-axis, must be rescaled to be between 0 and 1.  See \S \ref{sec:bounds} for a discussion of this.}}
   \label{fig:BPterms}
\end{figure}

Bernstein Polynomials (BPs) have a long history in mathematics and statistics since its original publication (Bernstein, 1912). Readers interested in knowing more about the use and origins of BPs in various scientific fields will find the article by \citet{Far12} very comprehensive and informative. Here we present a brief overview of the main properties and motivations of choosing BPs over other basis functions, such as splines or kernel methods. Given a continuous function $f(u)$ defined on the unit interval $[0, 1]$, the BP of degree $d$ corresponding to the function $f$ is defined as
\[B_d(u; f)=\sum_{k=0}^d f\left({k\over d}\right){d\choose k}u^k(1-u)^{d-k}\;\;\mbox{for}\;d=1,2,\ldots,\]
where $B_d(x,f)$ is a polynomial of degree at most $d${.  It} was shown by Bernstein that $\|B_d(\cdot,f)-f(\cdot)\|_\infty\equiv\sup_{u\in[0, 1]}|B_d(u,f)-f(u)|\rightarrow 0$ as $d\rightarrow\infty$, which constructively demonstrates that a continuous function on a closed interval can be uniformly approximated by a polynomial.
Figure \ref{fig:BPterms} provides plots of the basis function(s) with $d = 1, 2, 3, 5, 10, 20$.

The result extends to more than one dimension by taking the tensor product of the univariate BPs{.  In} this article we have used {suitable transformations} to approximate a bivariate density function defined on the unit square $[0, 1]^2$. In a statistical density estimation problem, we would not know the density function $f$ and we thus work with $\theta_k=f\left({k\over d}\right)$ for $k=0,1,\ldots,d$ and estimate the parameter vector $(\theta_0,\theta_1,\ldots,\theta_d)$ using likelihood and other standard statistical methods. The prime motivation for using BPs for estimating densities is that it not only enjoys many of the similar asymptotic properties as some of the other popular density estimation methods do (such as those based on kernels and B-splines), but it also has good asymptotic properties near boundaries. Its connection to {a} mixture of Beta densities {is described in \citet{Ten94}}. 
\citet{Babu06} provides much of the asymptotic properties of the BPs in multi-dimensions and show that densities can be estimated based on dependent sequence of multivariate random variables.
Moreover, it has been shown that BPs can also be used to make inference when the density or regression functions are subject {to shape constraints}
\citep[see][]{Tur14, Wan12}.

\section{More on the Bernstein polynomials\sout{s} model}
\label{sec:simulation}

\subsection{On choosing the upper and lower bounds for mass and radius: $\underline{M}$, $\overline{M}$, $\underline{R}$, $\overline{R}$.}\label{sec:bounds}

\begin{figure}[!h]
\centering
   \includegraphics[width=9cm]{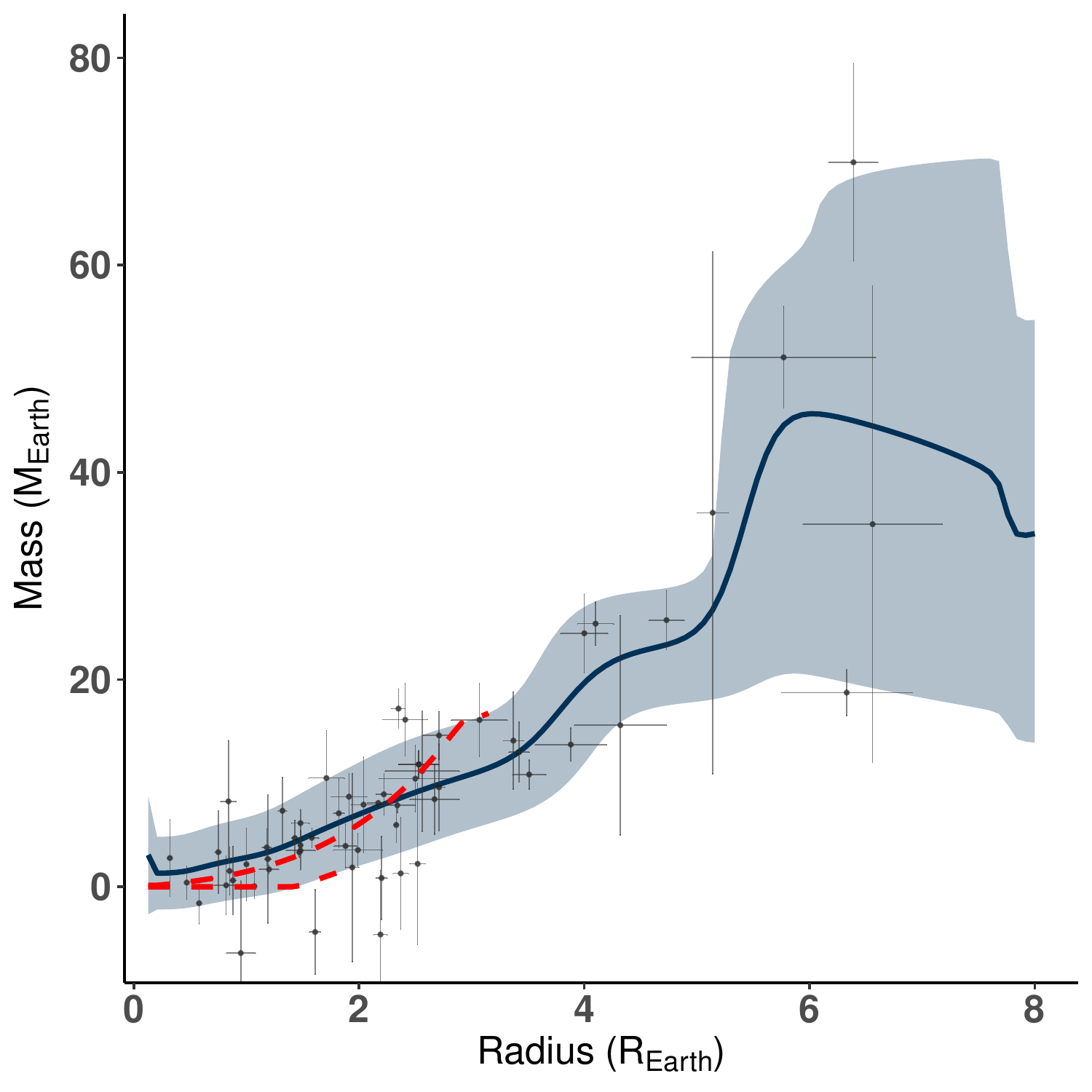} 
   \caption{
   \small Plot of the M-R relation with its prediction intervals{ when} $\overline{R}$ is chosen to be $8R_\oplus$ instead of $6.7R_\oplus$ ({as} in Figure \ref{fig:MRrelations}){. Comparing} with Figure \ref{fig:MRrelations}, {the} two M-R relations are similar in the region where $R \leq 6R_\oplus$, {where} a large amount of data is available. The current plot shows a {downward} trend for the M-R relation when $R > 6R_\oplus$ {because where} there is no data the model fits a line smoothly to the overall mean.
   The large prediction intervals show that this {downward} trend is highly {uncertain.  We note that in this plot, $\hat{d} = \hat{d}' = 35$, which was selected based on the cross-validation method described in \S \ref{sec:modinfer}.}
   }
   \label{fig:MRrelations-testbound}
\end{figure}

The Bernstein polynomials model requires {specifying} an upper bound and a lower bound for both mass and radius. 
Although in theory one could choose the upper bound and the lower bound to be any values{,
in practice
setting} the upper bounds and the lower bounds to different values{ will cause the results to} be different from the results in Figure \ref{fig:MRrelations} and Figure \ref{fig:keplerMRrelation}.
{In particular,} when there are no data in a region,  
the Bernstein polynomials model will fit a smooth line toward the overall mean. 
This happens when the upper bound is chosen {to be} too large.  {To illustrate this, we offer Fig \ref{fig:MRrelations-testbound} as a comparison to Figure \ref{fig:MRrelations}.}

In Figure \ref{fig:MRrelations-testbound}, we set the upper bound of masses to be $8 R_\oplus$, 
which is the same value used in WFR16's paper,
instead of $6.7 R_\oplus${. One can} immediately tell that the region between $6.7 R_\oplus$ and $8R_\oplus$ has a {downward} trend.
This is {because} the overall mean of {the dataset's planet masses} is smaller than the mass at $6.7R_\oplus$, which is where the last observation lies. 
{Because the model tends toward the overall mean when there is no data, a downward trend is produced.}
{This} trend might not provide a correct astrophysical interpretation of the M-R relation in this region{, and so we strongly caution against extrapolating this nonparametric relation}.
{That said,} the prediction intervals are very wide {in this region}, 
which suggests that {there is significant variability in the mean relation, which would be rectified with} more observations.

\subsection{On choosing the degrees $d$ and $d'$.}

\begin{figure}[!h]
\centering
   \includegraphics[width=10cm, angle=-90]{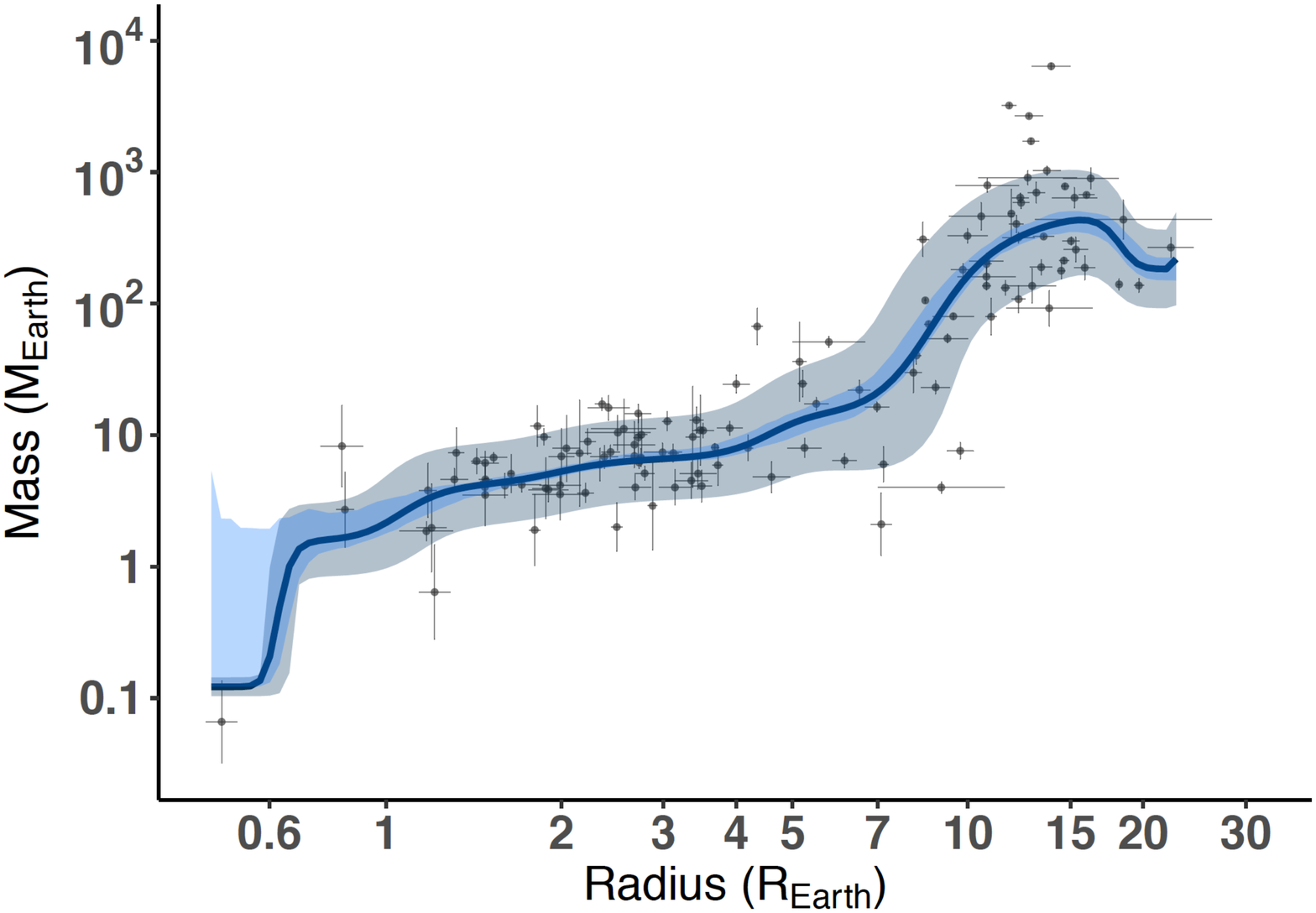}
   \caption{
   \small Plot of the M-R relation with its prediction intervals estimated {from} the \Kep data.
   We chose $\hat d = \hat d' = 55$, {where} the value{s} of $d$ and $d'$ are selected based on the cross-validation method 
   {discussed in \S \ref{sec:modinfer}.  Comparing} with Figure \ref{fig:keplerMRrelation}, {the} two M-R relations are similar.
   }
   \label{fig:keplerMRrelation-equal-degree}
\end{figure}

{Choosing} the degrees $d$ and $d'$ is {somewhat computationally expensive} due to the cross-validation method{'s searching for} the optimal $d$ and $d'$ from a large number of potential values. 
To {decrease} the computation time, 
one can force $d = d'$. 
We adopt this approach to choose $d$ and then re-estimate the parameters in the model.
Figure \ref{fig:MRrelations-testbound} and 
Figure \ref{fig:keplerMRrelation-equal-degree}
are the results produced by choosing $d = d'$.
{The} cross-validation method{ chooses} $d = d' = 35$ {for the first dataset} and $d = d' = 55${ for the second}.
Comparing these two figures to Figures \ref{fig:MRrelations}
and \ref{fig:keplerMRrelation} respectively, 
the results do not vary too much. 

\FloatBarrier
\section{A simulation study}
\label{sec:simulation}

\begin{figure*}
\centering
	\includegraphics[width=.4\textwidth]{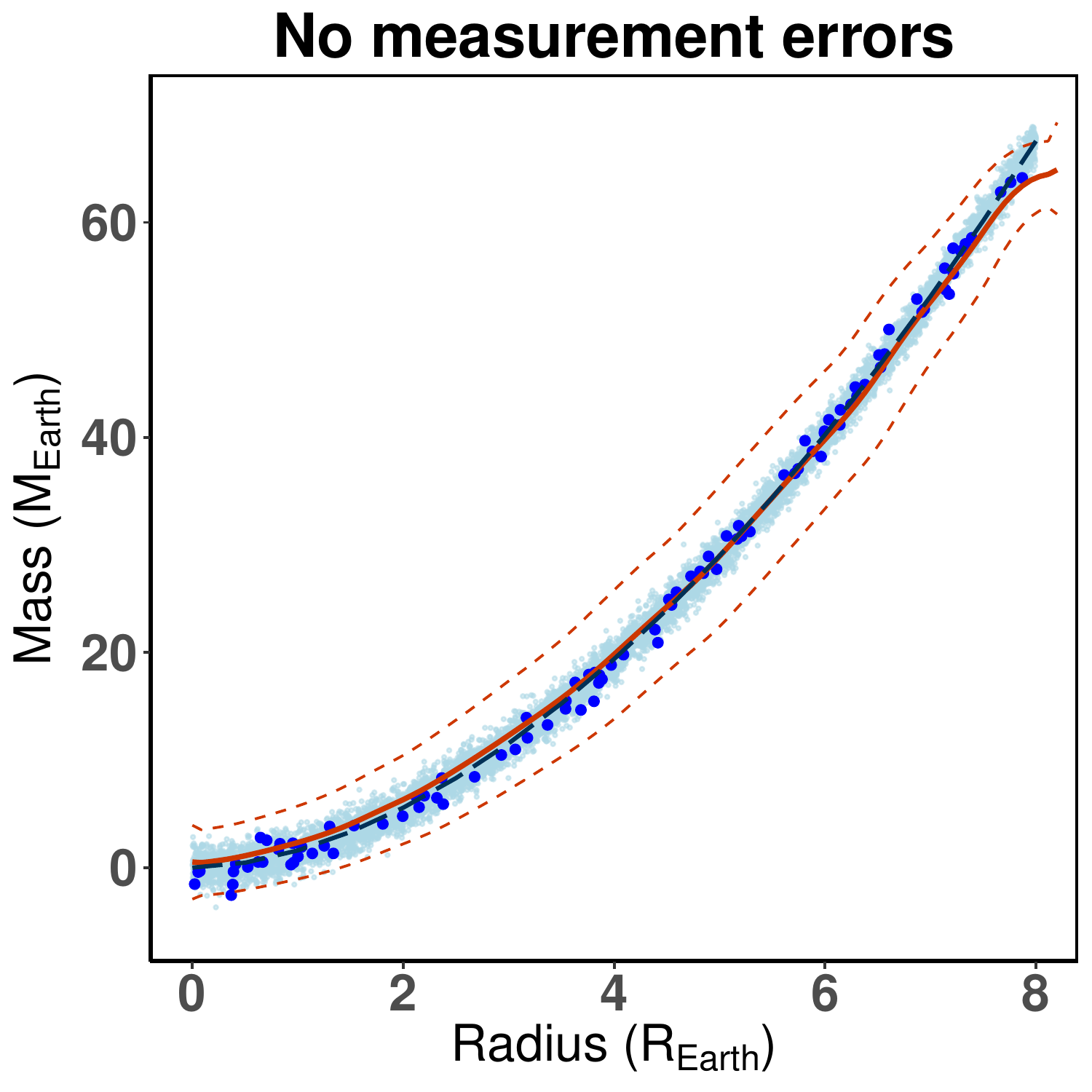} 
    \includegraphics[width=.4\textwidth]{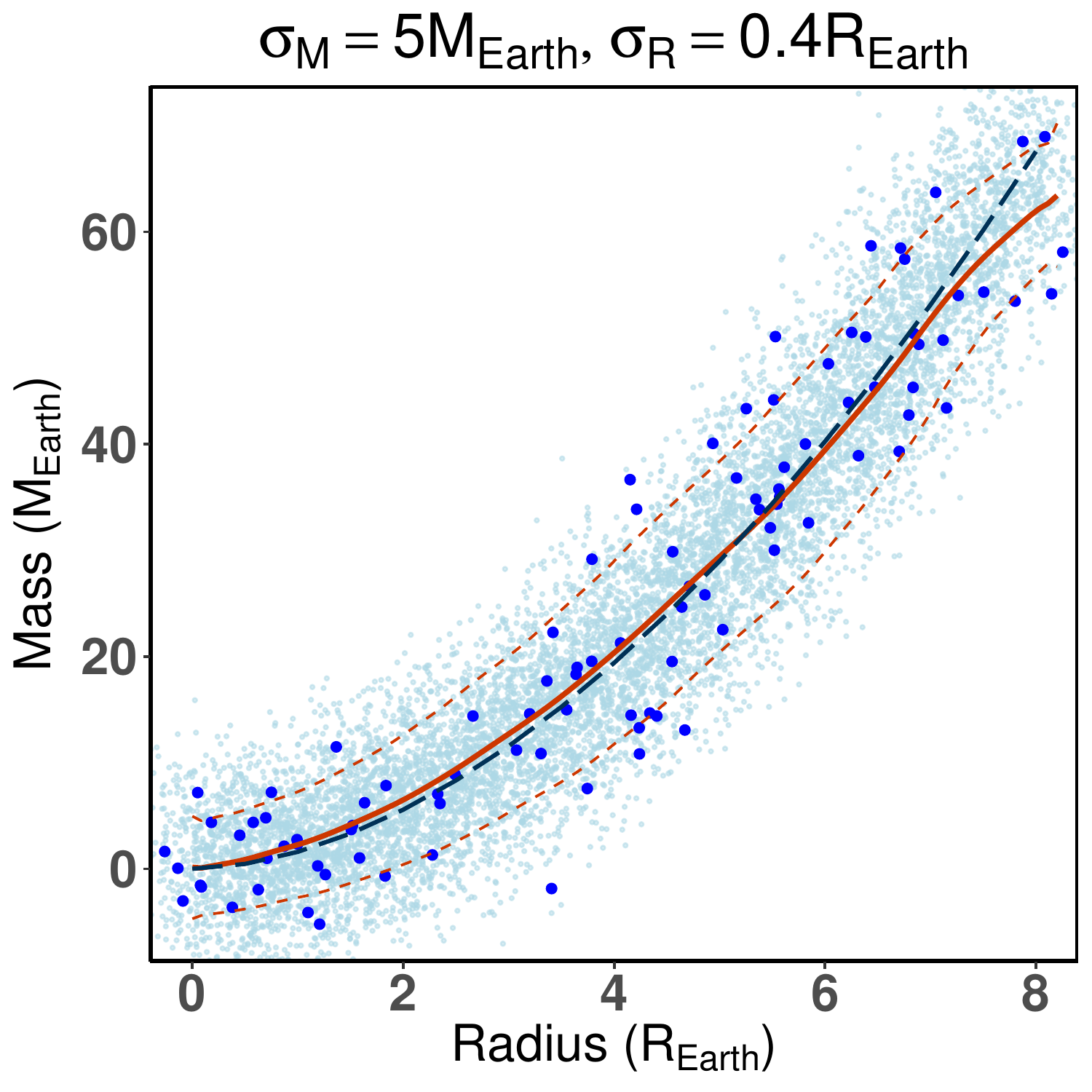}
    \includegraphics[width=.4\textwidth]{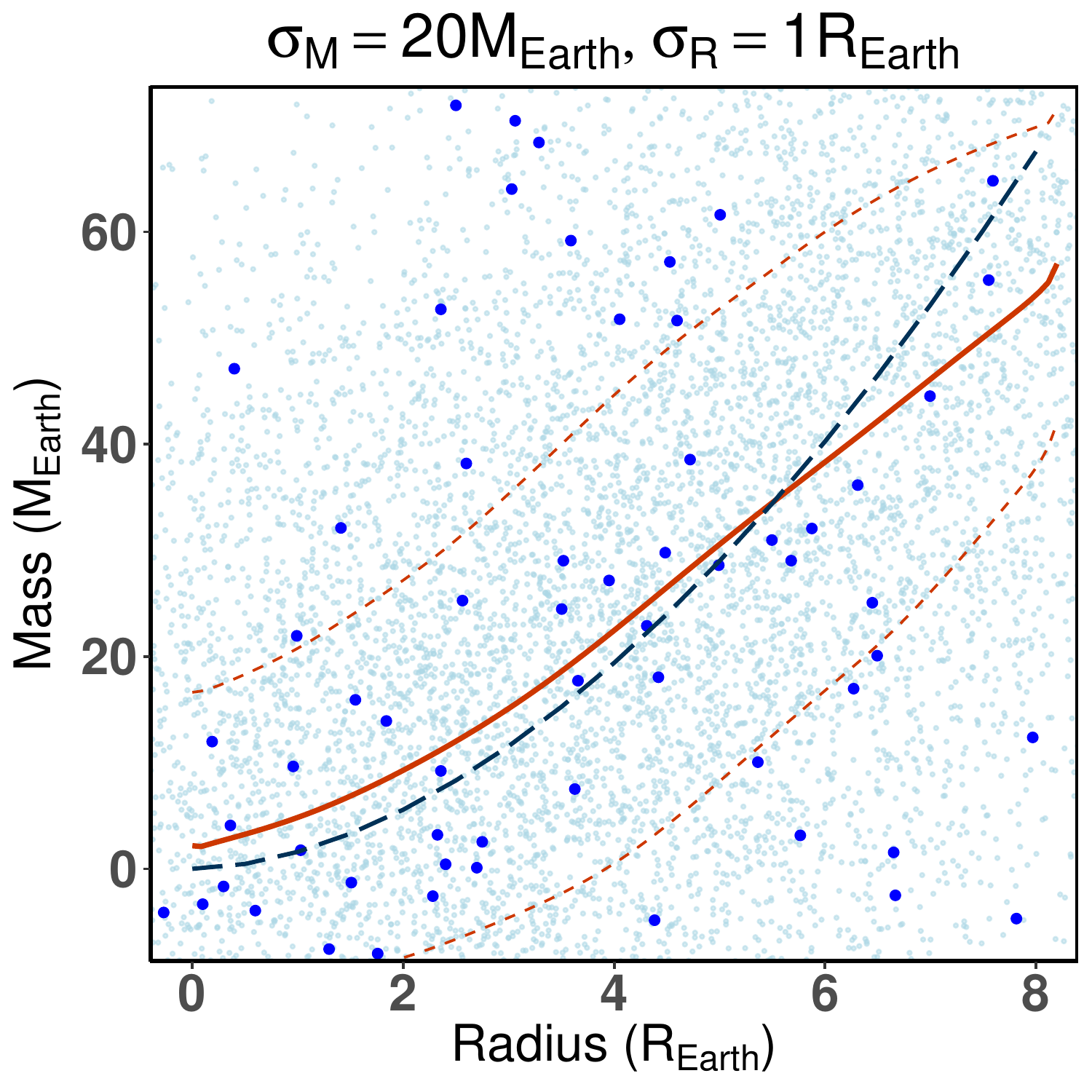}
   \caption{\small The performance of our nonparametric M-R relation as assessed with simulated data: only for the largest measurement uncertainties does the inferred M-R relation deviate from the truth.  Each simulation consists of 100 datasets, and each of those contain 100 data points; all 10,000 simulated planets are plotted as small light blue points, with the larger dark blue points {identifying} those from one randomly chosen dataset, to show the typical planet-to-planet variation in any individual dataset.  The different panels were generated with different measurement uncertainties; these are denoted above each panel in units of Earth mass and Earth radii.  For each panel, the black dashed line is the ``true'' M-R relation that we use to generate the individual dataset (see text for details).  The red lines illustrate the variability produced by fitting our model to each of the 100 different datasets: the solid red line denotes the mean of the most probable mass values (at each radius) across all 100 model fits (i.e.\ the mean of the 100 model fit means), and the dashed red lines denote the 16\% and 84\% confidence intervals of this most probable mass value (i.e.\ the standard error of the 100 model fit means).}
 \label{fig:simset}
\end{figure*}

To check how missing values and measurement errors could affect the estimation result, we conduct several simulation studies. 
We simulate data from a power-law relation and use the Bernstein polynomial model to estimate that function, {as} we are interest{ed in comparing} the M-R relations estimated under different settings.

In this simulation study, we define the ``true'' M-R relation as follows: $r \sim \mathcal{U}(0, 8)$, which stands for the uniform distribution between 0\Rearth and 8\Rearth, and $m \sim \mathcal{N}(1.6r^{1.8}, 1)$, i.e. the extended M-R relation found by WRF16 but with a smaller intrinsic scatter. 
We then perform a series of simulations where we generate 100 datasets, each of which has 100 observations.  What varies between the different simulations are the measurement errors: we choose the measurement errors on radii as 0.4, 1, 3 \Rearth and on masses as 5, 20, 35 \Mearth.
 
We fit each dataset to the model, choosing the degrees using the same 10-fold cross validation scheme that is described in \S \ref{sec:modinfer}.  As the measurement uncertainty increases, the observed data smear out the underlying M-R relation.  Information about the M-R relation is lost in this process, and so a less complex nonparametric model is required to fit the data.  This is reflected in a lower degree being recommended by our cross-validation scheme: for the original simulated dataset with no measurement errors, $d=90$, while $d=40,7,5$ 
for the {three} datasets with consecutively larger measurement errors.

In the following {figures}, we plot the estimated M-R relation (red line) along with the true M-R relation (black line){ and the}
16\% and 84\% confidence intervals. We also plot all 100 simulated datasets {in light blue in} the background and highlight one dataset from them {in dark blue}. 

{To summarize our findings, t}he first subplot in Figure \ref{fig:simset} suggests the nonparametric model did a good job {estimating the} M-R relation when the data is generated from the ``true'' model. 
When data have measurement errors, the quality of estimated M-R relation depends on the magnitude of measurement errors for a fixed sample size. The larger the measurement error{s} are, the worse the estimated M-R relations and the larger the confidence intervals. We also found {that} the true mean M-R relation is contained in the confidence intervals regardless {of} how large the measurement errors are.

\begin{figure}
\centering
\includegraphics[width=.4\textwidth]{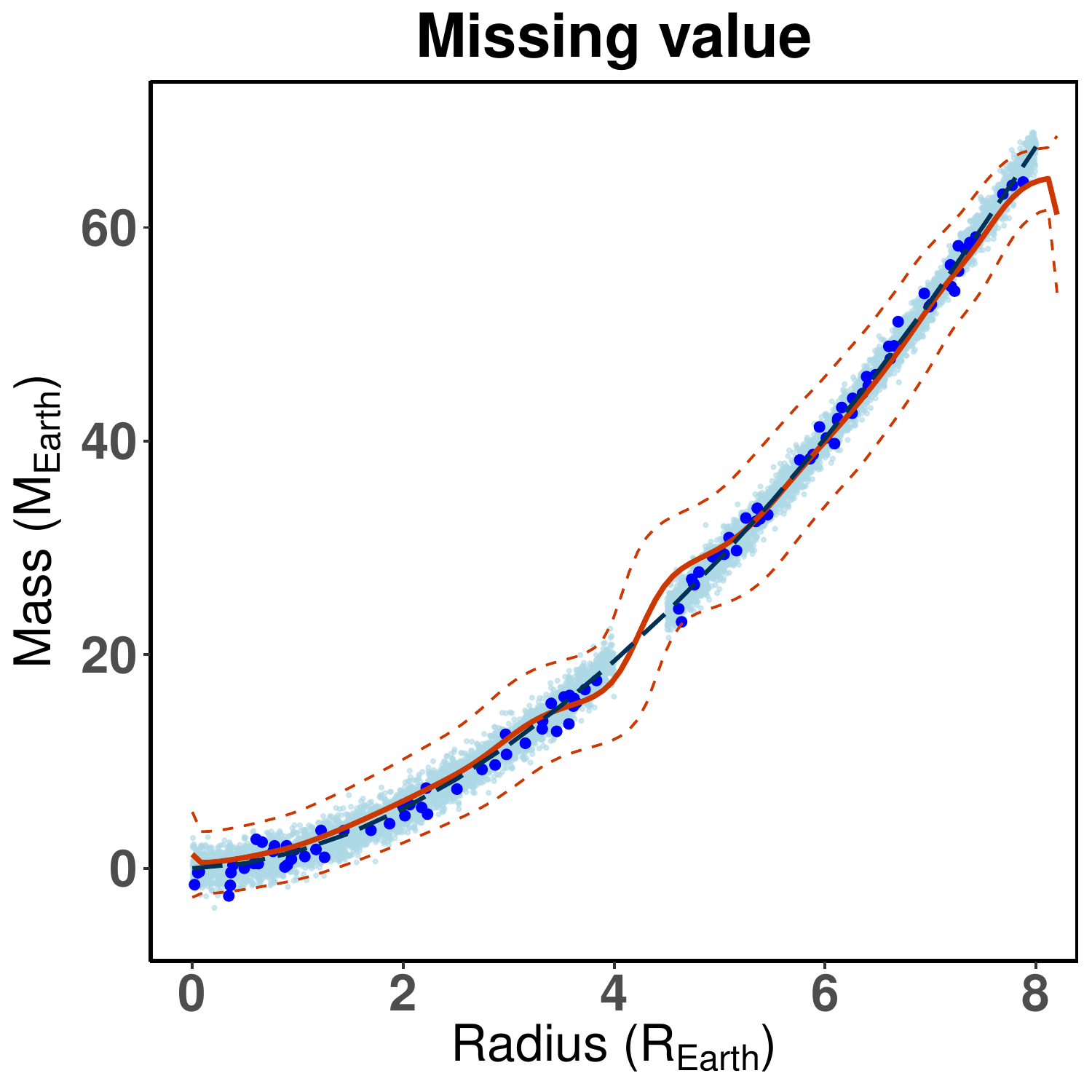}
\caption{The performance of our nonparametric method when the radius measurements are not distributed uniformly over $0<r<8$ \Rearth: the M-R relation defaults to a flat line where there is missing data.  The points and lines here show the same information as in Figure \ref{fig:simset}.} \label{fig:simmissdat}
\end{figure}

We also run a simulation study for a dataset with missing observations in some regions (non-uniform sampling). Figure \ref{fig:simmissdat} presents the estimation result when the data between 4.5 $R_\oplus$ and $5 R_\oplus$ are removed. Comparing to the first subplot in Figure \ref{fig:simset}, the M-R relation is different in the region where the data are missing. The model is trying to find the optimal polynomials to estimate the M-R relation. The polynomials may not necessarily to be linear, thus in the region with no data points, the M-R relation can be nonlinear. When there are missing values, Bernstein polynomials impute missing values by taking mean from the nearest region where data are presented. Thus the mean M-R relation is lower than the true relation on the left side of missing data region and higher than the true relation on the right side. While the mean M-R relation may depart from the true relation in the missing data region, the confidence intervals contains the true M-R relation, which does not rule out the possibility of obtaining the true relation.
This gives us confidence in applying our model to real \Kep and K2 data that have non-uniform sampling.



\begin{thebibliography}{}
\expandafter\ifx\csname natexlab\endcsname\relax\def\natexlab#1{#1}\fi

\end{thebibliography}


\begin{thebibliography}{}
\bibitem[Akeson et al.(2013)]{Ake13} Akeson, R.~L., Chen, X., Ciardi, D., et al.\ 2013, \pasp, 125, 989 

\bibitem[Babu \& Chaubey(2006)]{Babu06} Babu, G.~J., \& Chaubey, Y.~P. \ 2006, Statistics and Probability Letters, 76, 959
\bibitem[Babu et al.(2002)]{Babu02} Babu, G.~J., Canty, A.~J., \& Chaubey, Y.~P. \ 2002, Journal of Statistical Planning and Inference, 105, 377
\bibitem[Burt et al.(2018)]{Burt18} Burt, J., Holden, B.~P., Wolfgang, A., \& Bouma, L.~G.\ 2018, arXiv:1810.06569 

\bibitem[Chen \& Kipping(2017)]{Chen17} Chen, J., \& Kipping, D.\ 2017, \apj, 834, 17 
\bibitem[Coughlin et al.(2016)]{Cou16} Coughlin, J. L., Mullally, F., Thompson, S.~E., et al.\ 2016, \apjs, 224, 12 
\bibitem[Cumming et al.(2008)]{Cum08} Cumming, A., Butler, R.~P., Marcy, G.~W., et al.\ 2008, \pasp, 120, 531 

\bibitem[Dawson et al.(2015)]{Daw15} Dawson, R.~I., Chiang, E., \& Lee, E.~J.\ 2015, \mnras, 453, 1471
\bibitem[Dressing \& Charbonneau(2013)]{Dre13} Dressing, C.~D., \& Charbonneau, D.\ 2013, \apj, 767, 95 


\bibitem[Foreman-Mackey et al.(2014)]{FoM14} Foreman-Mackey, D., Hogg, D.~W., \& Morton, T.~D.\ 2014, \apj, 795, 64 
\bibitem[Fortney et al.(2007a)]{For07a} Fortney, J.~J., Marley, M.~S., \& Barnes, J.~W.\ 2007, \apj, 659, 1661 
\bibitem[Fortney et al.(2007b)]{For07b} Fortney, J.~J., Marley, M.~S., \& Barnes, J.~W.\ 2007, \apj, 668, 1267 
\bibitem[Fressin et al.(2013)]{Fre13} Fressin, F., Torres, G., Charbonneau, D., et al.\ 2013, \apj, 766, 81 
\bibitem[Fulton et al.(2017)]{Ful17} Fulton, B.~J., Petigura, E.~A., Howard, A.~W., et al.\ 2017, arXiv:1703.10375 
\bibitem[Farouki (2012)]{Far12} Farouki, R.~T., \ 2012, Computer Aided Geometric Design, 29, 379

\bibitem[Green(1995)]{Gre95} Green, P.~J.\ 1995, Biometrika, 82, 711

\bibitem[Hastie et al.(2001)]{Has01} Hastie, T., \& Tibshirani, R., \& Friedman J.~H.\ 2001, Springer
\bibitem[Hadden \& Lithwick(2014)]{Had14} Hadden, S., \& Lithwick, Y.\ 2014, Physics and Astronomy, 787, 80
\bibitem[Hadden \& Lithwick(2017)]{Had17} Hadden, S., \& Lithwick, Y.\ 2017, \aj, 154, 5 
\bibitem[Howard et al.(2010)]{How10} Howard, A.~W., Marcy, G.~W., Johnson, J.~A., et al.\ 2010, Science, 330, 653
\bibitem[Howard et al.(2012)]{How12} Howard, A.~W., Marcy, G.~W., Bryson, S.~T., et al.\ 2012, \apjs, 201, 15 
\bibitem[Hsu et al.(2018)]{Hsu18} Hsu, D.~C., Ford, E.~B., Ragozzine, D., \& Morehead, R.~C.\ 2018, arXiv:1803.10787

\bibitem[Ida \& Lin(2010)]{Ida10} Ida, S., \& Lin, D.~N.~C.\ 2010, \apj, 719, 810 
\bibitem[Isobe et al.(1990)]{Iso90} Isobe, T., Feigelson, E.~D., Akritas, M.~G., \& Babu, G.~J.\ 1990, \apj, 364, 104 

\bibitem[Jontof-Hutter et al.(2016)]{Jon16} Jontof-Hutter, D., Ford, E.~B., Rowe, J.~F., et al.\ 2016, \apj, 820, 39 

\bibitem[Kelly(2007)]{Kel07} Kelly, B.~C.\ 2007, \apj, 665, 1489 
\bibitem[Kruijer(2008)]{Kru08} Kruijer, W., \& van der Vaart, A.\ 2008, Journal of Statistical Planning and Inference, 138, 1981

\bibitem[Leblanc(2012)]{Leb12} Leblanc, A.\ 2012, Annals of the Institute of Statistical Mathematics, 919, 64 
\bibitem[Lee \& Chiang(2015)]{Lee15} Lee, E.~J., \& Chiang, E.\ 2015, \apj, 811, 41 
\bibitem[Lissauer et al.(2011)]{Lis11} Lissauer, J.~J., Ragozzine, D., Fabrycky, D.~C., et al.\ 2011, \apjs, 197, 8
\bibitem[Lopez \& Fortney(2014)]{Lop14} Lopez, E.~D., \& Fortney, J.~J.\ 2014, \apj, 792, 1
\bibitem[Lopez \& Rice(2016)]{Lop16} Lopez, E.~D., \& Rice, K.\ 2016, arXiv:1610.09390 

\bibitem[MacDonald et al.(2016)]{Mac16} MacDonald, M.~G., Ragozzine, D., Fabrycky, D.~C., et al.\ 2016, \aj, 152, 105 
\bibitem[Marcy et al.(2014)]{Mar14} Marcy, G.~W., Isaacson, H., Howard, A.~W., et al.\ 2014, \apjs, 210, 20
\bibitem[Mayor et al.(2011)]{May11} Mayor, M., Marmier, M., Lovis, C., et al.\ 2011, arXiv:1109.2497 
\bibitem[Mills et al.(2016)]{Mil16} Mills, S.~M., Fabrycky, D.~C., Migaszewski, C., et al.\ 2016, \nat, 533, 509 
\bibitem[Mills \& Mazeh(2017)]{Mil17} Mills, S.~M., \& Mazeh, T.\ 2017, \apjl, 839, L8 
\bibitem[Mordasini et al.(2012)]{Mor12} Mordasini, C., Alibert, Y., Georgy, C., et al.\ 2012, \aap, 547, A112 
\bibitem[Montet (2018)]{Mont18} Montet, B. \ 2018, \apj, 2, 28

\bibitem[Owen \& Wu(2017)]{Owen17} Owen, J.~E., \& Wu, Y.\ 2017, arXiv:1705.10810 

\bibitem[Petigura et al.(2013)]{Pet13} Petigura, E.~A., Marcy, G.~W., \& Howard, A.~W.\ 2013, \apj, 770, 69 
\bibitem[Petrone(1999a)]{Pet99a} Petrone, S.\ 1999, The Scandinavian Journal of Statistics, 26, 373
\bibitem[Petrone(1999b)]{Pet99b} Petrone, S.\ 1999, The Canadian Journal of Statistics, 27, 105
\bibitem[Petrone \& Wasserman(2002)]{Pet02} Petrone, S., \& Wasserman L.\ 2002, Journal of the Royal Statistical Society, Series B, 64, 79
\bibitem[Rogers et al.(2011)]{Rog11} Rogers, L.~A., Bodenheimer, P., Lissauer, J.~J., \& Seager, S.\ 2011, \apj, 738, 59

\bibitem[Rogers (2015)]{Rog15} Rogers, L.~A. \ 2015, \apj, 801, 41

\bibitem[Seager et al.(2007)]{Sea07} Seager, S., Kuchner, M., Hier-Majumder, C.~A., \& Militzer, B.\ 2007, \apj, 669, 1279 
\bibitem[Sestovic, Demory \& Queloz (2018)]{Ses18} Sestovic, M., Demory, B.-O., \& Queloz, D. \ 2018,
arXiv:1804.03075

\bibitem[Tenbusch(1994)]{Ten94} Tenbusch, A.\ 2014, Metrika, 41, 233
\bibitem[Turnbull \& Ghosh(2014)]{Tur14} Turnbull, B.~C., \& Ghosh, S.~K.\ 2014, Computational Statistics \& Data Analysis, 72, 13
\bibitem[Thorngren \& Fortney (2017)]{Tho17} Thorngren, D.~P., \& Fortney J.~J. \ 2017, arXiv:1709.04539

\bibitem[Valencia et al.(2007)]{Val07} Valencia, D., Sasselov, D.~D., \& O'Connell, R.~J.\ 2007, \apj, 665, 1413 

\bibitem[Wang \& Ghosh(2012)]{Wan12} Wang, J., \& Ghosh, S.~K.\ 2012, Computational Statistics \& Data Analysis, 56, 2729
\bibitem[Weiss \& Marcy(2014)]{Wei14} Weiss, L.~M., \& Marcy, G.~W.\ 2014, \apjl, 783, L6 (WM14)
\bibitem[Xie(2014)]{Xie14} Xie, J.~W.\ 2014, The Astrophysical Journal Supplement Series, 210, 25
\bibitem[Wolfgang \& Lopez(2015)]{Wol15} Wolfgang, A., \& Lopez, E.\ 2015, \apj, 806, 183 
\bibitem[Wolfgang et al.(2016)]{Wol16} Wolfgang, A., Rogers, L.~A., \& Ford, E.~B.\ 2016, \apj, 825, 19 (WRF16) 
\bibitem[Wu \& Lithwick(2013)]{WuY13} Wu, Y., \& Lithwick, Y.\ 2013, \apj, 772, 74

\bibitem[Youdin(2011)]{You11} Youdin, A.~N.\ 2011, \apj, 742, 38 
\bibitem[Zapolsky \& Salpeter(1969)]{Zap69} Zapolsky, H.~S., \& Salpeter, E.~E.\ 1969, \apj, 158, 809 
\end{thebibliography}


\end{document}